\documentclass[12pt,onecolumn]{article}

\usepackage{amsmath,amssymb,epsfig}
\newcommand{\figref}[1]{Fig.~\ref{fig:#1}}
\newcommand{\Real}{{\mathbb{R}}}
\newcommand{\Complex}{{\mathbb{C}}}

\global\def\putFrag#1#2#3#4{
\begin{figure}[tp]
\begin{center}
#4 \epsfxsize=#3in \epsfbox{#1.eps}
\end{center}
\caption{\small{#2}}
\label{fig:#1}
\end{figure}
}

\newtheorem{theorem}{Theorem}

\newtheorem{lemma}[theorem]{Lemma}

\newenvironment{proof}{{\sl Proof\/}:\ \ }

\begin{document}
\title{On the Optimality of the ARQ-DDF Protocol}
\author{Kambiz~Azarian, Hesham~El~Gamal, and
Philip~Schniter\footnote{The authors are with the Department of
Electrical and Computer Engineering, The Ohio State University,
Columbus, OH 43210 (e-mail: azariany@ece.osu.edu,
helgamal@ece.osu.edu, schniter@ece.osu.edu). This work was
supported in part by the National Science Foundation.}}

\maketitle

\begin{abstract}
The performance of the automatic repeat request-dynamic decode and
forward (ARQ-DDF) cooperation protocol is analyzed in two distinct
scenarios. The first scenario is the multiple access relay (MAR)
channel where a single relay is dedicated to simultaneously help
several multiple access users. For this setup, it is shown that the
ARQ-DDF protocol achieves the optimal diversity multiplexing
tradeoff (DMT) of the channel. The second scenario is the
cooperative vector multiple access (CVMA) channel where the users
cooperate in delivering their messages to a destination equipped
with multiple receiving antennas. For this setup, we develop a new
variant of the ARQ-DDF protocol where the users are purposefully
instructed not to cooperate in the first round of transmission.
Lower and upper bounds on the achievable DMT are then derived. These
bounds are shown to converge to the optimal tradeoff as the number
of transmission rounds increases.
\end{abstract}

\section{Background}
The dynamic decode and forward (DDF) protocol was proposed in
\cite{AES:05} as an efficient method to exploit cooperative
diversity in the half-duplex relay channel (the same protocol was
independently devised for other settings \cite{MOT:05,KS:05}). Here,
we combine the DDF protocol with the ARQ mechanism to derive new
variants that are matched to the multiple access relay (MAR) and
cooperative vector multiple access (CVMA) channels. These variants,
some of them presented in \cite{AES:04}, \cite{NAE:05} and
\cite{AE:061}, are shown to achieve the optimal tradeoff between
throughput and reliability, in the high signal to noise ratio (SNR)
regime. For simplicity of presentation, we restrict ourselves to the
two-user scenario. In principle, our results generalize to channels
with more users but the mathematical development becomes involved
and offers no additional insights.

Throughout the paper, all the channels are assumed to be flat
Rayleigh-fading and quasi-static. The quasi-static assumption
implies that the channel gains remain fixed over one coherence
interval and change, independently, from one coherence interval to
the next. In order to highlight the benefits of cooperation and ARQ,
as opposed to temporal interleaving, we adopt the long-term static
channel model of \cite{ECD:04} where all the ARQ rounds
corresponding to a certain message take place over the same
coherence interval. We further assume that the channel gains are
mutually independent with unit variance. The additive Gaussian noise
processes at different nodes are zero-mean, mutually-independent,
circularly-symmetric, and white. Furthermore, the variances of these
noise processes are proportional to one another such that there are
always \emph{fixed} offsets between different channels' SNRs. All
the nodes are assumed to operate synchronously and to have the same
power constraint. We impose a short-term power constraint on the
nodes ensuring that the average energy available for each symbol is
fixed. Except for section \ref{sec:CVMA}, where the destination is
assumed to have multiple receiving antennas, each node is equipped
with a single antenna. We adopt the coherent transmission paradigm
where only the receiving node of any link knows the channel gain.
Except for the ACK/NACK feedback bits, no other channel state
information (CSI) is available to the transmitting nodes. The
maximum number of transmission rounds is restricted to $L$, where
$L=1$ corresponds to the Non-ARQ scenario. We also assume the nodes
to operate in the half-duplex mode, i.e., at any point in time, a
node can either transmit or receive, but not both. This constraint
is motivated by the typically large difference between the incoming
and outgoing power levels. Throughout the paper, we use random
Gaussian code-books with asymptotically large block lengths to
derive information theoretic bounds on the achievable performance.
Results related to the design of practical coding/decoding schemes
that approach the fundamental limits established here will be
reported elsewhere (e.g., \cite{MAE:06}). Our analysis tool is the
diversity multiplexing tradeoff (DMT) introduced by Zheng and Tse
in~\cite{ZT:02}. To define DMT for a symmetric multiple access
channel with two users, we consider a family of codes
$C(\rho)=\{C_1(\rho),C_2(\rho)\}$ labeled by the operating SNR
$\rho$, such that the code $C_j(\rho)$ (used by user $j\in\{1,2\}$)
has a rate $R(\rho)/2$ bits per channel use (BPCU) and a maximum
likelihood (ML) error probability $P_{E_j}(\rho)$. For this family,
we define the multiplexing gain $r$ and diversity gain $d$ as
\begin{align}
&r \triangleq \lim_{\rho\rightarrow\infty}\frac{R(\rho)}{\log\rho} ,
&d \triangleq \min_{j \in \{1,2\}} \{-
\lim_{\rho\rightarrow\infty}\frac{\log P_{E_j}(\rho)}{\log\rho} \}.
\label{eq:3}
\end{align}
In ARQ scenarios, i.e., $L>1$, we replace $r$ with the effective
multiplexing gain $r_e$ introduced in \cite{ECD:04} to capture the
variable-rate nature of ARQ schemes.

The rest of this correspondence is organized as follows. Section
\ref{sec:MAR} is devoted to the MAR channel whereas our results for
the CVMA channel are presented in Section \ref{sec:CVMA}. We finish
with a few concluding remarks in Section~\ref{sec:conc}. To enhance
the flow of the paper, we only outline the main ideas of the proofs
in the body of the paper and relegate all the technical details to
the Appendix.

\section{The Multiple Access Relay Channel}\label{sec:MAR}
In the two-user multiple access relay (MAR) channel, one relay node
is assigned to assist the two multiple access users. The users are
not allowed to help each other (due to practical limitations, for
example). The relay node is constrained by the half-duplex
assumption. We proceed towards our main result in this section via a
step-by-step approach. First, we prove the optimality of the ARQ-DDF
protocol in the relay channel (i.e., a MAR channel with a single
user). The proof for this result introduces the machinery necessary
to handle the ARQ mechanism. In Lemma~\ref{thrm:2}, we analyze the
DDF protocol (without ARQ) in the two user MAR channel. This step
elucidates the multi-user aspects of the problem. Finally, we
combine these ideas in Theorem~\ref{thrm:3} to prove the optimality
of the DDF-ARQ protocol in the two-user MAR channel.
\begin{lemma} \label{thrm:1}
The optimal DMT for the relay channel with $L \geq 2$ ARQ rounds is
given by
\begin{align}
d_{\text{R}}(r_{\text{e}},L) = 2(1-\frac{r_{\text{e}}}{L})
\text{~~for~~} 1 > r_{\text{e}} \geq 0. \label{eq:7}
\end{align}
Furthermore, this optimal tradeoff is achieved by the proposed
ARQ-DDF protocol.
\end{lemma}
\begin{proof}(Sketch) The converse is readily obtained by applying
the results of \cite{ECD:04} to the genie-aided $2\times 1$
multiple-input multiple-output (MIMO) channel. The proof of the
achievability part, on the other hand, involves two main
ingredients; 1) showing that the average throughput, as a function
of SNR, converges asymptotically to the throughput corresponding to
one round of transmission, and 2) showing that the performance is
dominated by the errors for which $L$ rounds of transmission has
been requested. By combining these two steps, it can be shown that
from a probability of error perspective, the effective multiplexing
gain shrinks to $r_e/L$. The optimality of the ARQ-DDF protocol then
follows from the fact that the throughput range of interest (i.e.,
$r_e/L < 1/L$) falls within the optimality range of the DDF protocol
(i.e., $r \leq 1/2$) \cite{AES:05}.
\end{proof}

In our DDF protocol for the non-ARQ MAR channel, the two sources
transmit their individual messages during every symbol interval in
the codeword, while the relay listens to the sources until it
collects sufficient energy to decode \emph{both} of them error-free.
After decoding, the relay uses an \emph{independent} code book to
encode the two messages \emph{jointly}. The encoded symbols are then
transmitted for the rest of the codeword.
\begin{lemma} \label{thrm:2}
The optimal diversity gain for the symmetric two-user MAR channel is
upper bounded by
\begin{align}
d_{\text{MAR}}(r) &\leq \left\{
\begin{array}{lll}
2-r & \text{if} & \frac{1}{2} \geq r \geq 0 \\
3(1-r) & \text{if} & 1 \geq r \geq \frac{1}{2}
\end{array} \label{eq:8}
\right. .
\end{align}
Furthermore, the DMT achieved by the DDF protocol is lower bounded
by
\begin{align}
d_{\text{DDF-MAR}}(r) &\ge \left\{
\begin{array}{lll}
2-r & \text{if} & \frac{1}{2} \geq r \geq 0 \\
3(1-r) & \text{if} & \frac{2}{3} \geq r \geq \frac{1}{2}\\
2 \frac{1-r}{r} & \text{if} & 1 \geq r \geq \frac{2}{3}
\end{array} \label{eq:9}
\right. .
\end{align}
\end{lemma}
\begin{proof}(Sketch) The upper bound is obtained through a min-cut
max-flow argument. The proof for the achievability part follows the
same lines as that of Theorem $5$ in \cite{AES:05}.
\end{proof}
\putFrag{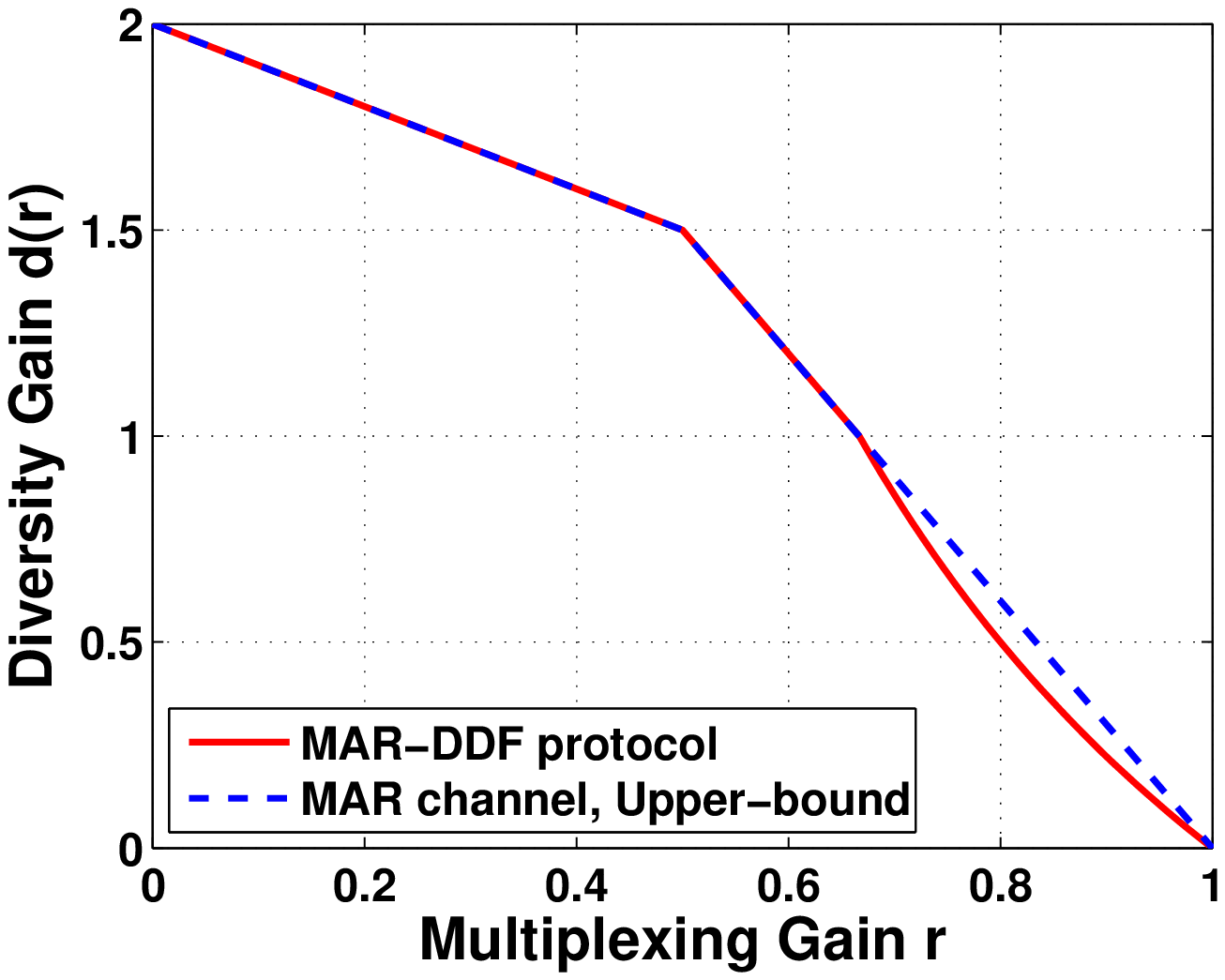}{The DMT achieved by the DDF protocol in the MAR
channel, along with an upper-bound on the achievable DMT
($L=1$).}{3.3}.{}

\figref{mar-ddf} compares the upper and lower bounds in
Lemma~\ref{thrm:2} where the optimality of the DDF protocol for $2/3
\geq r \geq 0$ is evident. This observation is the key to
establishing the following result.

\begin{theorem} \label{thrm:3}
The optimal DMT for the symmetric two-user MAR channel with $L \geq
2$ ARQ rounds is given by
\begin{align}
d_{\text{MAR}}(r_{\text{e}},L) = 2-\frac{r_{\text{e}}}{L}
\text{~~for~~} 1 > r_{\text{e}} \geq 0. \label{eq:10}
\end{align}
Furthermore, this optimal tradeoff is achieved by the proposed
ARQ-DDF protocol.
\end{theorem}
\begin{proof} (Sketch) The proof essentially reduces to combining
the results in Lemma \ref{thrm:1} and Lemma \ref{thrm:2}. One needs,
however, to be careful with a few technical details as reported in
the Appendix.
\end{proof}

We note that the optimality in Theorem~\ref{thrm:3} extends to the
$N$-user MAR channel. Overall, the main conclusion in this section
is establishing the fact that a \emph{single} relay can be
efficiently shared by \emph{several} multiple access users such that
it enhances the diversity gain achieved by \emph{all} of them.

\section{The Cooperative Vector Multiple-Access Channel} \label{sec:CVMA}
In the cooperative vector multiple access (CVMA) channel, the two
single-antenna users are allowed to assist each other, as long as
they do not violate the half duplex constraint. The challenge in
this scenario stems from the availability of two receiving antennas
at the destination, which increases the channel's degrees of freedom
to two. Loosely speaking, in order to exploit these two degrees of
freedom, the two users need to transmit \emph{new} independent
symbols continuously, which prevents them from cooperation under the
half duplex constraint (non-ARQ case). More rigorously, it is
straightforward to see that with $L=1$, any half-duplex cooperation
protocol that achieves full diversity (i.e., $d(0)=3$), falls short
of achieving full rate (i.e., $d(r)>0,$ for all $r<2$). To get
around this problem, in the case of $L \geq 2$, we purposefully
instruct the users \emph{not} to cooperate in the first round of
transmission. In fact, a user continues transmitting its message
while it receives NACK signals. Only when a user receives an ACK
signal, it starts listening to the other user (assuming the other
user has not been successfully decoded yet). Once the cooperating
user decodes the message of its partner, it starts helping (i.e.,
the typical DDF methodology). The following result establishes lower
and upper bounds on the DMT achieved by this protocol.

\begin{theorem} \label{thrm:4}
The optimal diversity gain for the symmetric two-user CVMA channel
with $L$ ARQ rounds is upper bounded by
\begin{align}
d_{CVMA}(r_{\text{e}},L) &\leq \min \left
\{3(1-\frac{r_{\text{e}}}{2L}), 4-\frac{3r_{\text{e}}}{L} \right \},
\text{~~for~~} 2 > r_{\text{e}} \geq 0. \label{eq:11}
\end{align}
For $L=2$, the diversity gain achieved by the proposed ARQ-DDF
protocol satisfies
\begin{align}
&d_{DDF-CVMA}(r_{\text{e}},2) \ge \left\{ \begin{array}{ll}
3-r_{\text{e}}, & 1 > r_{\text{e}} \ge 0 \\
4-2r_{\text{e}}, & \frac{4}{3} > r_{\text{e}} \ge 1 \\
2-\frac{r_{\text{e}}}{2}, & 2 > r_{\text{e}} \ge \frac{4}{3}
\end{array}\label{eq:12}
\right..
\end{align}
Furthermore, as $L$ increases, the ARQ-DDF diversity gain converges
to the optimal value, i.e.,
\begin{align}
\lim_{L \to \infty} d_{DDF-CVMA}(r_{\text{e}},L) = 3,\text{~~for~~}
2
> r_{\text{e}} \geq 0. \label{eq:121}
\end{align}
\end{theorem}
\begin{proof}(Sketch) The upper bound is obtained through a
min-cut max-flow argument. The lower bound, on the other hand,
results from analyzing a sub-optimal decoder which ignores parts of
the received signal to simplify the mathematical development. The
key observation for proving the asymptotic optimality result in this
theorem is that the dominant error event corresponds to the case
when one of the users is decoded successfully, while the other one
remains in error, even after $L$ rounds of transmission. In fact,
the proposed cooperation protocol is devised with this observation
in mind, i.e., the asymptotic optimality of the ARQ-DDF protocol
follows from the fact that, as $L$ increases, every user gets a
better chance of being helped by the other one.
\end{proof}
\putFrag{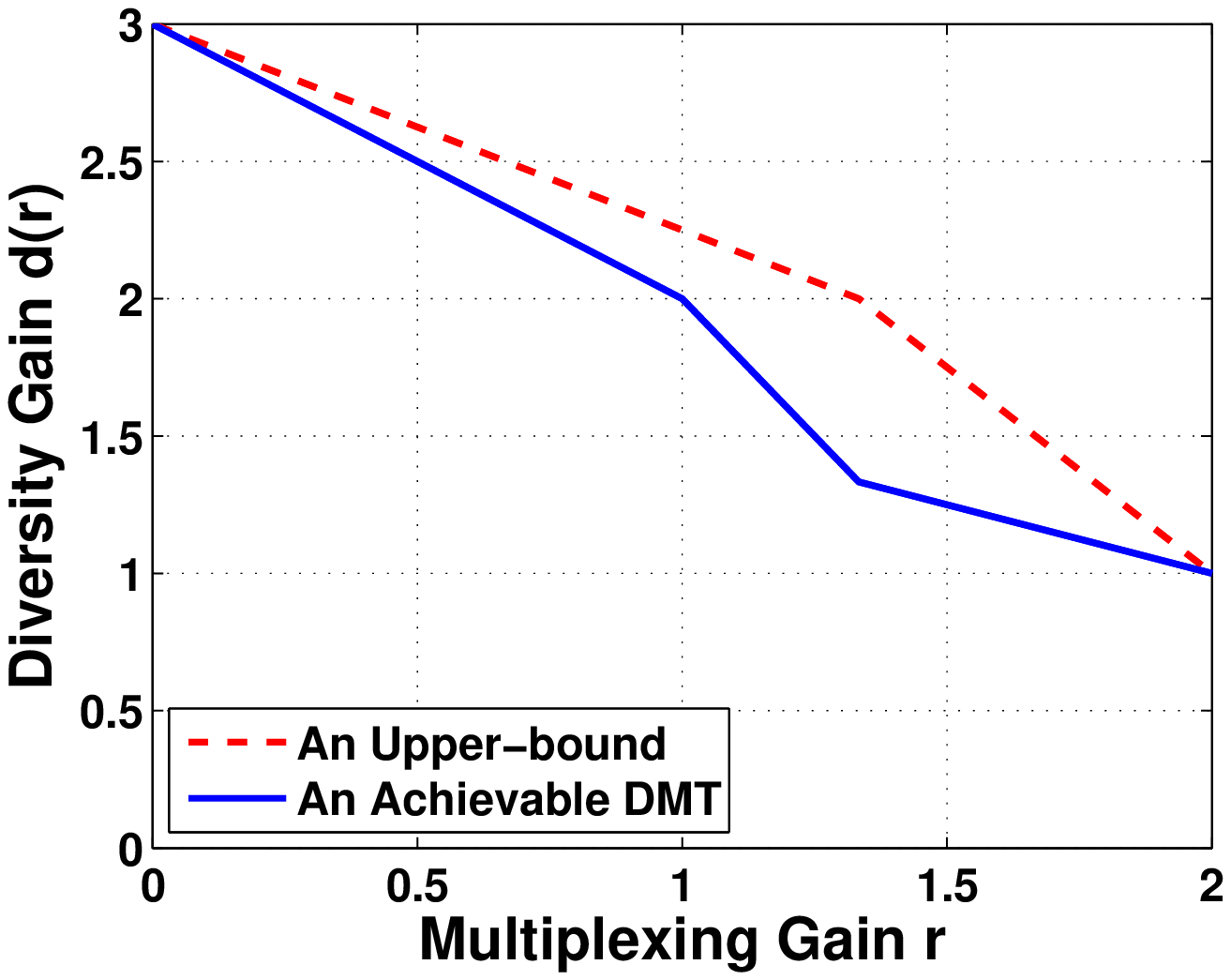}{ The DMT achieved by the ARQ-DDF protocol in the
CVMA channel, along with an upper-bound on the achievable DMT
($L=2$).}{3.3}{}

\figref{cma-ddf} compares the upper and lower bounds in
(\ref{eq:11}) and (\ref{eq:12}) for $L=2$. Clearly, this figure
shows the full diversity and full rate properties of the proposed
ARQ-DDF protocol. Finally, we observe that the analysis of the
ARQ-DDF protocol in Theorem~\ref{thrm:4} can be repeated for $L>2$.
We have, however, chosen to omit this analysis since it is rather
tedious and not necessarily inspiring.

\section{Conclusions}\label{sec:conc}
In this correspondence, we combined the DDF protocol with the ARQ
mechanism to develop efficient cooperation schemes for the MAR and
CVMA channels. The proposed ARQ-DDF protocol was shown to achieve
the optimal DMT for the MAR channel. In the CVMA scenario, we argued
that the ARQ-DDF protocol achieves significant cooperative diversity
gains while exploiting all of the channel's degrees of freedom and
despite the half duplex constraint. Furthermore, the diversity gain
achieved by the ARQ-DDF protocol in this scenario was shown to
converge to the optimal value, as the number of ARQ rounds
increases.

\section{Appendix}
We start with a few notations that are needed throughout the proofs.
The SNR of a link, $\rho$, is defined as
\begin{align}
\rho &\triangleq \frac{E}{\sigma^2}, \label{eq:1}
\end{align}
where $E$ denotes the average energy available for transmission of a
symbol across the link and $\sigma^2$ denotes the variance of the
noise observed at the receiving end of the link. We say that
$f(\rho)$ is \emph{exponentially} equal to $\rho^b$, denoted by
$f(\rho) \dot{=} \rho^b$, when
\begin{align}
\lim_{\rho \rightarrow \infty} \frac{\log f(\rho)}{\log \rho} &=
b. \label{eq:2}
\end{align}
In \eqref{eq:2}, $b$ is called the \emph{exponential order} of
$f(\rho)$. $\dot{\leq}$, $\dot{\geq}$ are defined similarly.

We denote the maximum allowable number of ARQ rounds by $L$ ($L=1$
corresponds to the non-ARQ scenario), where each round consists of
$T$ consecutive symbol intervals. We denote the first-round rate of
transmission by $R_1$ and the average throughput by $\eta$. These
two quantities are related through \cite{ECD:04}
\begin{align}
\eta &= \frac{R_1}{1+\sum_{\ell=1}^{L-1}p(\ell)}, \label{eq:39}
\end{align}
where $p(\ell)$ is the probability that the destination requests for
the $(\ell+1)$th round of transmission. We define the first-round
multiplexing gain, $r_1$, and effective multiplexing gain, $r_e$, as
\begin{align}
r_1 &\triangleq \lim_{\rho \to \infty} \frac{R_1}{\log \rho}, &r_e
\triangleq \lim_{\rho \to \infty} \frac{\eta}{\log \rho}.
\label{eq:4}
\end{align}
From \eqref{eq:39}, we note that if $\{p(\ell)\}_{\ell=1}^{L-1}$
decay polynomially with $\rho$, then $\eta \dot{=} R_1$ and thus
$r_e=r_1$.
\subsection{Proof of Lemma \ref{thrm:1}}
A simple min-cut max-flow examination reveals that the optimal
diversity gain for this channel is upper bounded by those of the $2
\times 1$ and $1 \times 2$ ARQ MIMO channels, thus
\begin{align}
d_{\text{R}}(r_e,L) &\leq 2(1-\frac{r_e}{L})\text{~~for~~} 1 > r_e
\geq 0. \label{eq:31}
\end{align}

Next we prove the achievability of this upper bound, by
characterizing the DMT for the proposed ARQ-DDF relay protocol. We
do this in two steps. First, we construct an ensemble of random
Gaussian codes and characterize its average error probability $P_E$,
and throughput $\eta$. We then show, through a simple expurgation
argument, that there are codes in the ensemble that perform at least
as well as these two averages, therefore achieving $P_E$ and $\eta$
\emph{simultaneously}.

Let $C(\rho) = \{C_s(\rho),C_r(\rho)\}$ denote the random codes used
by the source and the relay, respectively. These are codes of length
$LT$ symbols, rate $R_1/L$ BPCU and generated by an i.i.d complex
Gaussian random process of mean zero and variance $E$. Let us denote
the message to be sent by $m_0$. This message consists of $R_1T$
information bits. We denote the source and relay codewords
corresponding to $m_0$ by $\mathbf{x}_s(m_0)$ and
$\mathbf{x}_r(m_0)$, respectively. We also denote the signature of
message $m_0$ at the destination by $\mathbf{s}(m_0)$, i.e.,
\begin{align}
\mathbf{y} &= \mathbf{s}(m_0) + \mathbf{n}, \label{eq:32}
\end{align}
where $\mathbf{y}$ and $\mathbf{n}$ denote the destination received
signal and additive noise, respectively. It is important to realize
that $\mathbf{s}(m_0)$, not only depends on the message $m_0$, but
also on the channel realization and the relay noise. Also, notice
that $\mathbf{x}_r(m_0)$ is only partially transmitted. This is
because the half-duplex relay, itself, needs first to listen to the
source to be able to decode the message. Finally, we use superscript
$\ell$ to denote the portion of the signal that corresponds to the
first $\ell$ rounds of transmission. The decoder
$\{\mathbf{\varphi},\mathbf{\psi}\}$, consists of two functions,
$\mathbf{\varphi} = \{\varphi^{\ell}\}_{\ell=1}^L$ and
$\mathbf{\psi} = \{\psi^{\ell}\}_{\ell=1}^{L-1}$.
\begin{itemize}
\item At round $\ell$ ($L \geq \ell \geq 1$), $\varphi^{\ell}$ outputs the
  message that minimizes
  $|\mathbf{y}^{\ell}-\mathbf{s}^{\ell}|^2$, i.e.
  \begin{align}
  \varphi^{\ell}(\mathbf{y}^{\ell}) &= \arg \min_{m}
  |\mathbf{y}^{\ell}-\mathbf{s}^{\ell}(m)|^2,\text{~~for~~} L \geq \ell \geq
  1. \label{eq:33}
  \end{align}
  We denote the event that $\varphi^{\ell}(\mathbf{y}^{\ell})$ differs from
  $m_0$, with $m_0$ denoting the transmitted message, by $E^{\ell}$.
\item At round $\ell$ ($L-1 \geq \ell \geq 1$),
  $\psi^{\ell}$ outputs a one, if $m$ is the
  \emph{unique} message for which
  \begin{align}
  |\mathbf{y}^{\ell}-\mathbf{s}^{\ell}(m)|^2 &\leq \ell T (1+\delta)
   \sigma^2,\label{eq:37}
  \end{align}
  where $\sigma^2$ denotes the destination noise variance and $\delta$ is
  some positive value. In any other case, $\psi^{\ell}$
  outputs a zero. We denote the event that $\psi^{\ell}$
  outputs a one, by $A^{\ell}$.
\end{itemize}
The decoder uses $\mathbf{\varphi}$ and $\mathbf{\psi}$ to decode the
message as follows:
\begin{enumerate}
\item At the end of round $\ell$, ($ L-1 \geq \ell \geq 1$), the decoder
  computes both, $\varphi^{\ell}(\mathbf{y}^{\ell})$ and
  $\psi^{\ell}(\mathbf{y}^{\ell})$. If $\psi^{\ell}(\mathbf{y}^{\ell})=1$,
  then the decoder declares $\varphi^{\ell}(\mathbf{y}^{\ell})$ as the
  received message and sends back an ACK. Otherwise, it requests for another
  round of transmission by sending back a NACK signal.
\item At the end of the $L$th round, though, the decoder outputs
  $\varphi^{L}(\mathbf{y}^{L})$ as the received message.
\end{enumerate}
To characterize the average error probability, $P_E$, we first use
the Bayes' rule to write
\begin{align}
P_E &\leq P_{E|\overline{E}_r} + P_{E_r}, \nonumber
\end{align}
where $E_r$ and $\overline{E}_r$ denote the events that the relay
makes an error in decoding the message, and its complement,
respectively. Since the relay starts transmission only after the mutual
information between its received signal and the source signal
exceeds $R_1T$, we have (e.g., refer to Theorem $10.1.1$ in
\cite{CT:91})
\begin{align}
P_{E_r} &\leq \epsilon, \text{~~for any~~}\epsilon >0.\nonumber
\end{align}
This means that
\begin{align}
P_E &\dot{\leq} P_{E|\overline{E}_r}. \nonumber
\end{align}
For the sake of notational simplicity, in the sequel, we denote
$P_{E|\overline{E}_r}$ by $P_E$. In characterizing $P_E$, we take
the approach of El Gamal, Caire and Damon in \cite{ECD:04}, i.e., we
upper bound $P_E$ by
\begin{align}
P_E \leq \sum_{\ell=1}^{L-1} P_{E^{\ell},A^{\ell}} + P_{E^{L}}.
\label{eq:34}
\end{align}
Notice that $P_{E^{\ell},A^{\ell}}$ upper bounds the probability of
\emph{undetected} errors with $\ell < L$ rounds of transmission,
while $P_{E^L}$ upper bounds the probability of \emph{decoding}
errors at the end of $L$ transmission rounds. The next step in
characterizing $P_E$ is to show that, for the decoder of interest,
the undetected errors do not dominate the overall error event, i.e.,
\begin{align}
P_E &\dot{\leq} P_{E^L}. \label{eq:35}
\end{align}
Toward this end, we note that
\begin{align}
P_{E^{\ell},A^{\ell}} &\leq \Pr\{|\mathbf{n}^{\ell}|^2 > \ell
T(1+\delta) \sigma^2\}. \label{eq:36}
\end{align}
To understand \eqref{eq:36}, let us assume that
$|\mathbf{n}^{\ell}|^2 \leq \ell T(1+\delta)\sigma^2$. One can then use
\eqref{eq:32} to conclude that
$|\mathbf{y}^{\ell}-\mathbf{s}^{\ell}(m_0)|^2 \leq \ell
T(1+\delta)\sigma^2$, where $m_0$ denotes the transmitted message. This,
however, is in contradiction with $E^{\ell} \cap A^{\ell}$. This is
because the latter event implies that some message $m_1$, other than $m_0$,
is the \emph{unique} message for which
$|\mathbf{y}^{\ell}-\mathbf{s}^{\ell}(m_1)|^2 \leq \ell
T(1+\delta)\sigma^2$. Thus $E^{\ell} \cap A^{\ell}
\subseteq\{|\mathbf{n}^{\ell}|^2 > \ell T(1+\delta) \sigma^2\}$, which
means that \eqref{eq:36} is indeed true. Now, $|\mathbf{n}^{\ell}|^2$ has a
central Chi-squared distribution with $2 \ell T$ degrees of freedom. One
can use the Chernoff bound to upper bound the tail of this distribution to
get
\begin{align}
\Pr\{|\mathbf{n}^{\ell}|^2 > \ell T(1+\delta) \sigma^2\} &\leq
(1+\delta)^{\ell T} e^{-\ell T \delta}. \label{eq:24}
\end{align}
This, however, in conjunction with \eqref{eq:36} means that, for any $\delta >
0$, it is possible to choose $T$ large enough, such that
\begin{align}
P_{E^{\ell},A^{\ell}} \leq \epsilon,\text{~~for any~~}\epsilon >
0.\label{eq:55}
\end{align}
Now, \eqref{eq:35}  follows from \eqref{eq:55}, together with
\eqref{eq:34}. Examination of $E^L$ reveals that $P_{E^L}$ is the
probability of error for the DDF relay protocol at a multiplexing
gain of $r_1/L$, i.e.
\begin{align}
P_{E^L} &\dot{=} \rho^{-d_{\text{DDF-R}}(\frac{r_1}{L})},\label{eq:41}
\end{align}
where $d_{\text{DDF-R}}(\cdot)$ denotes the diversity gain achieved
by the DDF relay protocol \cite{AES:05}. Using \eqref{eq:35}, we
conclude
\begin{align}
P_E &\dot{\leq} \rho^{-d_{\text{DDF-R}}(\frac{r_1}{L})}.\label{eq:38}
\end{align}
The final step is to show that for $1 > r_1 \geq 0$, we have
$r_1=r_e$. Towards this end, we use \eqref{eq:39} and notice that
for the scenario of interest,
\begin{align}
p(\ell) &\triangleq
P_{\overline{A^{1}},\cdots,\overline{A^{\ell}}},\text{~~for~~} L >
\ell > 0, \label{eq:40}
\end{align}
where $\overline{A^{\ell}}$ denotes the complement of $A^{\ell}$. To
characterize $p(\ell)$, we first upper bound it by
\begin{align}
p(\ell) &\leq P_{\overline{A^{\ell}}}.\label{eq:45}
\end{align}
Careful examination of $\overline{A^{\ell}}$ reveals that
\begin{align}
\overline{A^{\ell}} &= R_0^{\ell} \cup R_1^{\ell}, \label{eq:93}
\end{align}
where $R_0^{\ell}$ denotes the subset of destination signals,
$\mathbf{y}^{\ell}$, not included in any of the spheres of squared
radius $\ell T(1+\delta) \sigma^2$, centered at the signatures of
all possible messages $\{\mathbf{s}^{\ell}(m)\}$, i.e.,
\begin{align}
R_0^{\ell} &\triangleq \cap_{m} \{| \mathbf{y}^{\ell} -
\mathbf{s}^{\ell}(m) |^2 > \ell T(1+\delta)\sigma^2 \}. \label{eq:42}
\end{align}
$R_1^{\ell}$, on the other hand, represents the subset of destination
signals, $\mathbf{y}^{\ell}$, included in more than one such spheres, i.e.,
\begin{align}
R_1^{\ell} &\triangleq \cup_{m} R_1^{\ell}(m),\label{eq:43}
\end{align}
where
\begin{align}
R_1^{\ell}(m) &\triangleq \cup_{m_1 \neq m}
\{|\mathbf{y}^{\ell}-\mathbf{s}^{\ell}(m) |^2 \leq \ell T(1+\delta)\sigma^2,
|\mathbf{y}^{\ell}-\mathbf{s}^{\ell}(m_1)|^2 \leq \ell
T(1+\delta)\sigma^2\}.\label{eq:51}
\end{align}
Note that $R_1^{\ell}(m)$ consists of the intersections of the sphere
corresponding to message $m$, with those of the other messages. Next, we
assume that message $m_0$ is transmitted and characterize
$P_{\overline{A^{\ell}}|m_0}$. Toward this end, we write
\begin{align}
\overline{A^{\ell}} &= R_0^{\ell} \cup R_1^{\ell},\text{~~or}\nonumber\\
\overline{A^{\ell}} &= R_0^{\ell} \cup \Big(R_1^{\ell} \cap
\overline{R_1^{\ell}(m_0)}\Big) \cup R_1^{\ell}(m_0),\label{eq:56}
\end{align}
where the last step follows from \eqref{eq:43}. Now, examining
\eqref{eq:42}, we have
\begin{align}
R_0^{\ell} &\subseteq \{|\mathbf{y}^{\ell}-\mathbf{s}^{\ell}(m_0) |^2>\ell
T(1+\delta)\sigma^2\}. \nonumber
\end{align}
On the other hand, realizing that $R_1^{\ell} \cap
\overline{R_1^{\ell}(m_0)}$ consists of the intersections of all spheres
excluding the one corresponding to $m_0$, gives
\begin{align}
R_1^{\ell} \cap \overline{R_1^{\ell}(m_0)} &\subseteq
\{|\mathbf{y}^{\ell}-\mathbf{s}^{\ell}(m_0) |^2>\ell
T(1+\delta)\sigma^2\}. \nonumber
\end{align}
Thus
\begin{align}
R_0^{\ell} \cup \Big(R_1^{\ell} \cap \overline{R_1^{\ell}(m_0)}\Big) &\subseteq
\{|\mathbf{y}^{\ell}-\mathbf{s}^{\ell}(m_0) |^2>\ell
T(1+\delta)\sigma^2\}, \nonumber\\
&= \{|\mathbf{n}^{\ell}|^2>\ell T(1+\delta)\sigma^2\},\nonumber
\end{align}
where the last step follows from \eqref{eq:32} and the assumption that
$m_0$ is the transmitted message. Recalling \eqref{eq:24}, we conclude
\begin{align}
\Pr\{R_0^{\ell} \cup \big(R_1^{\ell} \cap \overline{R_1^{\ell}(m_0)}\big)|m_0\}
&\leq \epsilon,\text{~~for any~~}\epsilon > 0.\nonumber
\end{align}
This, together with \eqref{eq:56}, means that
\begin{align}
P_{\overline{A^{\ell}}|m_0} &\dot{\leq}
\Pr\{R_1^{\ell}(m_0)|m_0\}.\label{eq:50}
\end{align}
Characterization of $\Pr\{R_1^{\ell}(m_0)|m_0\}$ is a little bit more
involved. In particular, if we let $a \triangleq
\mathbf{s}^{\ell}(m_0)-\mathbf{s}^{\ell}(m_1)$, $b \triangleq
\mathbf{y}^{\ell}-\mathbf{s}^{\ell}(m_0)$ and $\Delta \triangleq \ell
T(1+\delta)\sigma^2$, then
\begin{align}
\{|b|^2 \leq \Delta, |a+b|^2 \leq \Delta \} =& \{|a|^2 \leq 4\Delta, |b|^2
\leq \Delta, |a+b|^2 \leq \Delta \} \cup \nonumber \\
&\{|a|^2 > 4\Delta, |b|^2 \leq \Delta, |a+b|^2 \leq \Delta \}.\nonumber
\end{align}
Since the second set on the right-hand side of this expression is empty, we
get
\begin{align}
\{|b|^2 \leq \Delta, |a+b|^2 \leq \Delta \} &= \{|a|^2 \leq 4\Delta, |b|^2
\leq \Delta, |a+b|^2 \leq \Delta \} \nonumber\\
&\subseteq \{|a|^2 \leq 4\Delta\}\nonumber.
\end{align}
This, along with \eqref{eq:51}, results in
\begin{align}
R_1^{\ell}(m_0) &\subseteq
\cup_{m_1 \neq m_0}
\{|\frac{\mathbf{s}^{\ell}(m_0)-\mathbf{s}^{\ell}(m_1)}{2}|^2 \leq \ell
T(1+\delta)\sigma^2\}, \nonumber
\end{align}
which gives
\begin{align}
\Pr\{R_1^{\ell}(m_0)|m_0\} &\leq \sum_{m_1 \neq m_0}
\Pr\{|\frac{\mathbf{s}^{\ell}(m_0)-\mathbf{s}^{\ell}(m_1)}{2}|^2 \leq \ell
T(1+\delta)\sigma^2\}.\label{eq:47}
\end{align}
We identify the right hand side of \eqref{eq:47}, as the union bound
on the ML error probability, conditioned on transmission of $m_0$
(refer to equation $(17)$ in \cite{ZT:02}, for a very similar
expression). Here, the noise variance is $(1+\delta)\sigma^2$, the
code length is $\ell T$ and the rate is $R_1/\ell$ BPCU. As a result
\begin{align}
\Pr\{R_1^{\ell}(m_0)\} &\dot{\leq}
\rho^{-d_{\text{DDF-R}}(\frac{r_1}{\ell})},\text{~~for~~} L> \ell >
0,\label{eq:52}
\end{align}
where $d_{\text{DDF-R}}(.)$ is the diversity gain achieved by the
DDF relay protocol \cite{AES:05}. Now \eqref{eq:52}, along with
\eqref{eq:50} and \eqref{eq:45} gives
\begin{align}
p(\ell) &\dot{\leq}
\rho^{-d_{\text{DDF-R}}(\frac{r_1}{\ell})},\text{~~for~~} L > \ell >
0.\label{eq:59}
\end{align}
This means that over the range of $1>r_1 \geq 0$, the probabilities
$\{p(\ell)\}_{\ell=1}^{L-1}$, decay polynomially with $\rho$. As a result,
over this range, $r_e$ equals $r_1$, i.e.
\begin{align}
r_e &=r_1,\text{~~for~~} 1 > r_1 \geq 0.\label{eq:57}
\end{align}
Now \eqref{eq:57}, together with \eqref{eq:38}, and the fact that
for $1 > r_e \geq 0$,
$d_{\text{DDF-R}}(\frac{r_e}{L})=2(1-\frac{r_e}{L})$, gives
\begin{align}
P_E &\dot{\leq} \rho^{-2(1-\frac{r_e}{L})},\text{~~for~~} 1 >
r_e \geq 0. \label{eq:58}
\end{align}
Note that \eqref{eq:58} only characterizes the relation between the
\emph{average} error probability $P_E$ and \emph{average} throughput
$\eta$. To complete the proof, we need to show that there exists a
code in the ensemble, that \emph{simultaneously} achieves $P_E$ and
$\eta$, as characterized by \eqref{eq:58}. Toward this end, we use
Lemma $11$ of \cite{ECD:04}. The application of this lemma to the
case of interest is immediate and thus the proof is complete.

\subsection{Proof of Lemma \ref{thrm:2}}
A simple max-cut min-flow examination reveals that the optimal
diversity gain for this channel is upper bounded by
\begin{align}
d_{\text{MAR}}(r) &\leq \min \{d_{3 \times 1}(r), d_{2 \times 2}(r),
d_{2 \times 1}(\frac{r}{2}), d_{1 \times 2}(\frac{r}{2})\},
\label{eq:60}
\end{align}
where $d_{m \times n}(\cdot)$ denotes the optimal diversity gain for
an $m \times n$ MIMO channel. Now, \eqref{eq:60} results in
\eqref{eq:8} and the proof of the converse part is complete.

In order to derive a lower bound on the diversity gain achieved by
the DDF MAR protocol, we upper bound the \emph{source-specific} ML
error probabilities, with that of the \emph{joint} ML decoder.
Furthermore, instead of characterizing the latter probability for
specific codes, in the sequel, we characterize its average, $P_E$,
over the ensemble of random Gaussian codes. It is then
straightforward to see that there exists a code in the ensemble,
whose error probability is better than $P_E$. To characterize $P_E$,
we use the Bayes' rule to derive the following upper bound
\begin{align}
P_E &\leq P_{E|\overline{E}_r} + P_{E_r}, \nonumber
\end{align}
where $E_r$ and $\overline{E}_r$ denote the events that the relay
makes errors in decoding the messages, and its complement,
respectively. Next, we note that if we denote the signals
transmitted by the two sources and the relay by
$\{x_{j,k}\}_{k=1}^T$ and $\{x_{r,k}\}_{k=T'+1}^{T}$ , respectively,
and the signals received by the relay and the destination by
$\{y_{r,k}\}_{k=1}^{T'}$ and $\{y_k\}_{k=1}^T$, then the number of
symbol intervals $T'$ that the relay waits before decoding the
messages satisfies
\begin{align}
\frac{TR}{2} &\leq
I(\{x_{1,k}\}_{k=1}^{T'};\{y_{r,k}\}_{k=1}^{T'}|\{x_{2,k}\}_{k=1}^{T'}),
\label{eq:13}\\
\frac{TR}{2} &\leq
I(\{x_{2,k}\}_{k=1}^{T'};\{y_{r,k}\}_{k=1}^{T'}|\{x_{1,k}\}_{k=1}^{T'}),
\label{eq:14}\\
TR &\leq
I(\{x_{1,k}\}_{k=1}^{T'},\{x_{2,k}\}_{k=1}^{T'};\{y_{r,k}\}_{k=1}^{T'}).
\label{eq:15}
\end{align}
In these expressions, $R$ is the \emph{total} data rate (in BPCU) at
the destination and $I(.;.)$ denotes the mutual information
function. Now, observing that \eqref{eq:13}, \eqref{eq:14} and
\eqref{eq:15} guarantee that (e.g., refer to section $14.3.1$ of
\cite{CT:91})
\begin{align}
P_{E_r} &\leq \epsilon, \text{~~for any~~} \epsilon > 0, \nonumber
\end{align}
we conclude
\begin{align}
P_E &\dot{\leq} P_{E|\overline{E}_r}. \nonumber
\end{align}
For the sake of notational simplicity, in the sequel,
$P_{E|\overline{E}_r}$ is denoted by $P_E$. In characterizing $P_E$,
we follow the approach of Tse, Viswanath and Zheng \cite{TVZ:03}, by
partitioning the error event $E$ into the set of partial error
events $E_I$, i.e.,
\begin{align}
E &= \bigcup_I E_I. \nonumber
\end{align}
where $I$ denotes any \emph{nonempty} subset of $\{1,2\}$ and $E_I$
(referred to as type-$I$ error) is the event that the joint ML
decoder incorrectly decodes the messages from sources whose indices
belong to $I$ while correctly decoding all other messages. Because
the partial error events are mutually exclusive, we have
\begin{align}
P_E &= \sum_I P_{E_I}. \label{eq:16}
\end{align}
To characterize $P_{E_I}$, we use the Bayes' rule to derive the
following upper-bound
\begin{align}
P_{E_I} &= P_{O_I}P_{E_I|O_I} +
P_{E_I,\overline{O}_I} \nonumber \\
P_{E_I} &\leq P_{O_I}+P_{E_I,\overline{O}_I} , \nonumber
\end{align}
where $O_I$ and $\overline{O}_I$ denote the type-$I$ outage event and its
complement, respectively. The type-$I$ outage event is defined such
that $P_{O_I}$ dominates $P_{E_I,\overline{O}_I}$, i.e.
\begin{align}
P_{E_I,\overline{O}_I} &\dot{\leq} P_{O_I}. \label{eq:17}
\end{align}
Thus,
\begin{align}
P_{E_I} &\dot{\leq} P_{O_I}. \label{eq:18}
\end{align}
Characterization of $O_I$, however, requires derivation of
$P_{PE_I|g,h}$, i.e., the joint ML decoder's type-$I$ pairwise error
probability (PEP), conditioned on a particular channel realization
and averaged over the ensemble of random Gaussian codes. For this
purpose, let us denote the gain of the channels connecting the two
sources to the relay by $h_1$ and $h_2$ and those of the channels
connecting the two sources and the relay to the destination by
$g_1$, $g_2$ and $g_r$, respectively. It is then straightforward to see
that (refer to \cite{AES:05})
\begin{align}
P_{PE_{\{1\}}|g,h} &\leq
(1+\frac{1}{2}\rho|g_1|^2)^{-T'}(1+\frac{1}{2}\rho(|g_1|^2
+|g_r|^2))^{-(T-T')}, \nonumber\\
P_{PE_{\{1,2\}}|g,h} &\leq
(1+\frac{1}{2}\rho(|g_1|^2+|g_2|^2))^{-T'}
(1+\frac{1}{2}\rho(|g_1|^2+|g_2|^2+|g_r|^2))^{-(T-T')}.
\nonumber
\end{align}
Notice that since the channel is symmetric, $P_{E_{\{1\}}} \dot{=}
P_{E_{\{2\}}}$, i.e., we do not need to characterize
$P_{PE_{\{2\}}|g,h}$. Let us denote the exponential orders of
$\{1/|g_j|^2\}_{j=1}^2$ and $1/|g_r|^2$ by $\{v_j\}_{j=1}^2$ and $v_r$,
respectively, and those of $\{1/|h_j|^2\}_{j=1}^2$ by $\{u_j\}_{j=1}^2$.
Realizing that, at a rate of $R=r \log \rho$ BPCU and a codeword length of
$T$, there are a total of $\rho^{Tr}$ codewords in the code, we get
\begin{align}
P_{E_{\{1\}}|v,u} &\dot{\leq}
\rho^{-T[f(1-v_1)^++(1-f)(1-\min{\{v_1,v_r\}})^+ - \frac{r}{2}]},
\label{eq:19}\\
P_{E_{\{1,2\}}|v,u} &\dot{\leq}
\rho^{-T[f(1-\min{\{v_1,v_2\}})^++(1-f)(1-\min{\{v_1,v_2,v_r\}})^+-r]},
\label{eq:20}
\end{align}
where $f \triangleq T'/T$. Careful examination of \eqref{eq:19}
reveals that defining $O_{\{1\}}^+$ as
\begin{align}
O_{\{1\}}^{+} &\triangleq \{(v_1,\cdots,u_2) \in \Real^{5+} |
f(1-v_1)^++(1-f)(1-\min\{v_1,v_r\})^+ \leq \frac{r}{2} \},
\label{eq:21}
\end{align}
satisfies \eqref{eq:17}. In this expression, $O_{\{1\}}^+$ denotes
$O_{\{1\}} \cap \Real^{5+}$, where $\Real^{n+}$ represents the set
of all nonnegative real $n$-tuples (for an explanation on why we are
only concerned about $O_I^+$, please refer to \cite{AES:05}). This
is because, if \eqref{eq:21} is satisfied, then through choosing a
large enough $T$, $P_{E_{\{1\}},\overline{O}_{\{1\}}}$ can be made
arbitrarily small. Likewise, defining $O_{\{1,2\}}^+$ as
\begin{align}
&O_{\{1,2\}}^{+} \triangleq \nonumber \\
&\{(v_1,\cdots,u_2) \in \Real^{5+} |
f(1-\min\{v_1,v_2\})^++(1-f)(1-\min\{v_1,v_2,v_r\})^+ \leq r \},
\label{eq:22}
\end{align}
satisfies \eqref{eq:17}. With the type-$I$ outage events specified,
the only thing left is to characterize $P_{O_I}$ (please refer to
\cite{AES:05}), i.e.
\begin{align}
P_{O_I} &\dot{=} \rho^{-d_I(r)} \text{~~where~~} d_I(r) \triangleq
\inf_{O_I^+} \{v_1+v_2+v_r+u_1+u_2\}. \label{eq:23}
\end{align}
Toward this end, we use \eqref{eq:21} to derive $\inf_{(v_1,v_2,v_r)
\in O_{\{1\}}^+} \{v_1+v_2+v_r\}$, as a \emph{function} of $f$, i.e.
\begin{align}
\inf_{(v_1,v_2,v_r) \in O_{\{1\}}^+} \{v_1+v_2+v_r\}(f) &=
\lambda_{\{1\}}(f), \nonumber
\end{align}
where
\begin{align}
\lambda_{\{1\}}(f) &\triangleq \left\{
\begin{array}{ll}
2-r, & \frac{1}{2} > f \ge 0\\
2 - \frac{r}{2(1-f)}, & 1-\frac{r}{2} > f \ge \frac{1}{2} \\
\frac{2-r}{2f}, & 1 \geq f \ge 1-\frac{r}{2}
\end{array}\label{eq:26}
\right.
\end{align}
Likewise, one can use \eqref{eq:22} to derive
\begin{align}
\inf_{(v_1,v_2,v_r) \in O_{\{1,2\}}^+} \{v_1+v_2+v_r\}(f) &=
\lambda_{\{1,2\}}(f), \nonumber
\end{align}
where
\begin{align}
\lambda_{\{1,2\}}(f) &\triangleq \left\{
\begin{array}{ll}
3(1-r), & \frac{2}{3} > f \ge 0\\
3-\frac{r}{1-f}, & 1-r > f \ge \frac{2}{3}\\
2 \frac{1-r}{f}, & 1 \geq f \ge 1-r
\end{array}, \text{~~if~~} \frac{1}{3} > r \geq 0 \label{eq:133}
\right.
\end{align}
or
\begin{align}
\lambda_{\{1,2\}}(f) &\triangleq \left\{
\begin{array}{ll}
3(1-r), & \frac{2}{3} > f \ge 0\\
2 \frac{1-r}{f}, & 1 \geq f \ge \frac{2}{3}
\end{array},\text{~~if~~} 1 \geq r \geq \frac{1}{3} \label{eq:27}
\right.
\end{align}
Now, to complete the derivation of $d_I(r)$, we need to characterize
$\inf_{(u_1,u_2) \in \Real^{2+}} \{u_1+u_2\}(f)$. For this purpose,
we use \eqref{eq:13}, \eqref{eq:14} and \eqref{eq:15} to derive
\begin{align}
T' &= \min \{ T, \max \{ \lceil\frac{TR}{2\log_2{(1+\min \{ |h_1|^2,
|h_2|^2\}c\rho)}}\rceil, \lceil
\frac{TR}{\log_2{(1+(|h_1|^2+|h_2|^2)c\rho)} } \rceil \} \},
\nonumber
\end{align}
where $c$ is the ratio of destination noise variance to that of the
relay and $\lceil x \rceil$ denotes the closest integer to $x$
towards plus infinity. In terms of channel exponential orders, this
last expression can be rewritten as
\begin{align}
f=& \min{\{1,
\max{\{\frac{r}{2(1-\max{\{u_1,u_2\}})^+},\frac{r}{(1-\min{\{u_1,u_2\}})^+}\}}\}},
& (u_1,u_2) \in \Real^{2+}. \label{eq:25}
\end{align}
Now, using \eqref{eq:25}, one can show that
\begin{align}
\inf_{(u_1,u_2) \in \Real^{2+}} \{u_1+u_2\}(f) &=
\lambda(f), \nonumber
\end{align}
where
\begin{align}
\lambda(f) &\triangleq \left\{
\begin{array}{ll}
2 (1-\frac{r}{f}), & \frac{3r}{2} > f \ge r\\
1-\frac{r}{2f}, & 1 \geq f \ge \frac{3r}{2}
\end{array}\label{eq:28}
\right. .
\end{align}
Using \eqref{eq:23} we conclude
\begin{align}
d_{\{1\}}(r) &= \inf_{f} \{\lambda(f) + \lambda_{\{1\}}(f)\}, \nonumber
\end{align}
which can be characterized using \eqref{eq:26} and \eqref{eq:28},
\begin{align}
d_{\{1\}}(r) &= \left\{
\begin{array}{ll}
2-r, & \frac{1}{2} > r \ge 0\\
\frac{4-5r}{2(1-r)}, & \frac{2}{3} \geq r \ge \frac{1}{2}\\
\frac{2-r}{2r}, & 1 \geq r \ge \frac{2}{3}
\end{array}\label{eq:29}
\right. .
\end{align}
Similarly,
\begin{align}
d_{\{1,2\}}(r) &= \inf_{f} \{\lambda(f) + \lambda_{\{1,2\}}(f)\},
\nonumber
\end{align}
which can be derived using \eqref{eq:133}, \eqref{eq:27} and
\eqref{eq:28},
\begin{align}
d_{\{1,2\}}(r) &= \left\{
\begin{array}{ll}
3(1-r), & \frac{2}{3} > r \ge 0\\
2\frac{1-r}{r}, & 1 \geq r \ge \frac{2}{3}
\end{array}\label{eq:30}
\right. .
\end{align}
Now, \eqref{eq:29} and \eqref{eq:30}, together with \eqref{eq:23},
\eqref{eq:18} and \eqref{eq:16}, result in \eqref{eq:9} and thus
complete the proof of the achievability part.

\subsection{Proof of Theorem \ref{thrm:3}}
A simple max-cut min-flow examination reveals that the optimal
diversity gain for this channel is upper bounded by
\begin{align}
d_{\text{MAR}}(r_e,L) &\leq \min \{d_{3 \times 1}(r_e,L), d_{2
\times
  2}(r_e,L), d_{2 \times 1}(\frac{r_e}{2},L), d_{1 \times
  2}(\frac{r_e}{2},L)\}, \label{eq:61}
\end{align}
where $d_{m \times n}(\cdot,\cdot)$ denotes the optimal diversity
gain for an $m \times n$ ARQ MIMO channel (refer to \cite{ECD:04}).
Now, \eqref{eq:61} results in
\begin{align}
d_{\text{MAR}}(r_e,L) &\leq 2-\frac{r_e}{L},\text{~~for~~} 1 > r_e
\geq 0, \label{eq:62}
\end{align}
which completes the proof of the converse.

Next, we prove that the proposed protocol achieves this upper bound.
To do this, we only need to describe the encoder and the decoder.
The rest of the proof then follows that of Lemma \ref{thrm:1}, line
by line. Toward this end, let
$C(\rho)=\{C_1(\rho),C_2(\rho),C_r(\rho)\}$ denote the random codes
used by the two sources and the relay, respectively. These are codes
of length $LT$, generated by an i.i.d complex Gaussian random
process of mean zero and variance $E$. The rates of these codes are
different, though. While $C_1(\rho)$ and $C_2(\rho)$ are of rate
$R_1/2L$ BPCU, $C_r(\rho)$ is of rate $R_1/L$ BPCU. In other words,
the relay code has twice the rate of the source codes. This means
that, corresponding to each pair of source codewords
$(\mathbf{x}_1(m_1),\mathbf{x_2}(m_2)) \in C_1(\rho) \times
C_2(\rho)$, there exists a codeword $\mathbf{x}_r(\mathbf{m}) \in
C_r(\rho)$, where $\mathbf{m} \triangleq (m_1,m_2)$. We call
$\mathbf{m}$ the joint message. Note that since each of the two
source messages, i.e. $m_1$ and $m_2$, consists of $R_1T/2$
information bits, the joint message $\mathbf{m}$, has a total of
$R_1T$ information bits in it. As before, we denote the destination
signature corresponding to the joint message $\mathbf{m}$, by
$\mathbf{s}(\mathbf{m})$, i.e.,
\begin{align}
\mathbf{y} &= \mathbf{s}(\mathbf{m})+\mathbf{n}.\nonumber
\end{align}
In order to decode the source messages and produce the ACK/NACK
signals, the destination uses a \emph{joint} bounded distance
decoder. This decoder is identical to the one devised for the
ARQ-DDF relay protocol (refer to Lemma \ref{thrm:1}), with the only
modification that the joint message $\mathbf{m}$ takes the role of
$m$ everywhere, e.g., $\varphi^{\ell}(\cdot)$ is now defined as
(compare to \eqref{eq:33})
\begin{align}
\varphi^{\ell}(\mathbf{y}^{\ell}) &\triangleq \arg \min_{\mathbf{m}}
|\mathbf{y}^{\ell}-\mathbf{s}^{\ell}(\mathbf{m})|^2,\text{~~for~~}L \geq
\ell \geq 1.\nonumber
\end{align}
In the proposed decoder, the destination provides a total of one bit
of feedback, for the two sources, per transmission round. Therefore,
there is no need for defining \emph{source-specific}
$\mathbf{\varphi}(\cdot)$ and $\mathbf{\psi}(\cdot)$ functions. With
the encoder and decoder defined, one can now follow the same steps
taken in the proof of Lemma \ref{thrm:1} to show that (compare to
\eqref{eq:38})
\begin{align}
P_E &\dot{\leq} \rho^{-d_{\text{DDF-MAR}}(\frac{r_1}{L})},\label{eq:63}
\end{align}
and that (compare to \eqref{eq:59})
\begin{align}
p(\ell) &\dot{\leq}
\rho^{-d_{\text{DDF-MAR}}(\frac{r_1}{\ell})},\text{~~for~~} L > \ell >
0.\label{eq:64}
\end{align}
Now, \eqref{eq:63} and \eqref{eq:64}, together with the fact that
for $1 > r_e \geq 0$, $d_{\text{DDF-MAR}}(\frac{r_e}{L})=
2-\frac{r_e}{L}$ (refer to \eqref{eq:9}), result in (compare to
\eqref{eq:58})
\begin{align}
P_E &\dot{\leq} \rho^{-(2-\frac{r_e}{L})},\text{~~for~~}1 > r_e \geq
0.\label{eq:65}
\end{align}
It is then straightforward to use Lemma $11$ of \cite{ECD:04} to
show that there are codes in the ensemble $\{C(\rho)\}$, that
achieve \eqref{eq:65}. This proves that the upper bound
\eqref{eq:62} is achievable and thus, completes the proof.

\subsection{Proof of Theorem \ref{thrm:4}}
A simple min-cut max-flow examination reveals that the optimal
diversity gain for this channel is upper bounded by
\begin{align}
d_{\text{CVMA}}(r_e,L) &\leq \min \{d_{2 \times 2}(r_e,L), d_{1
\times 3}(\frac{r_e}{2},L)\}, \label{eq:66}
\end{align}
where $d_{m \times n}(\cdot,\cdot)$ denotes the optimal diversity
gain for an $m \times n$ ARQ MIMO channel. Now, \eqref{eq:66}
results in \eqref{eq:11} and the proof of the converse part is
complete.

To prove the achievability part, let
$C(\rho)=\{C_1(\rho),C_2(\rho)\}$ denote the random codes used by
the two sources. These are codes of length $2T$ symbols, rate
$R_1/4$ BPCU, and generated by an i.i.d complex Gaussian random
process of mean zero and variance $E$. Let us also denote the two
messages to be sent by $m_1$ and $m_2$. Note that each message
consists of $R_1T/2$ information bits, such that the joint message
$\mathbf{m}\triangleq (m_1,m_2)$ consists of a total of $R_1T$ bits.
We denote the codewords corresponding to $m_1$ and $m_2$ by
$\mathbf{x}_1(m_1)$ and $\mathbf{x}_2(m_2)$. As before, the
destination signatures of $m_1$, $m_2$ and $\mathbf{m}$ are denoted
by $\mathbf{S}(m_1)$, $\mathbf{S}(m_2)$ and
$\mathbf{S}(\mathbf{m})$, respectively. Thus
\begin{align}
\mathbf{Y} &= \mathbf{S}(\mathbf{m})+\mathbf{N},\text{~~or}\nonumber\\
\mathbf{Y} &= \mathbf{S}(m_1)+\mathbf{S}(m_2)+\mathbf{N},\nonumber
\end{align}
where $\mathbf{Y} \in \Complex^{2 \times 2T}$ and $\mathbf{N} \in
\Complex^{2 \times 2T}$ represent the destination received signal
and additive noise, respectively. We denote the signal received
through antenna $j \in \{1,2\}$ by $\mathbf{y}_j$. Similarly, the
contribution of message $m_i$ $i \in \{1,2\}$, to the signal
received through antenna $j$, is denoted by $\mathbf{s}_j(m_i)$. As
before, we use the superscript $\ell$ to denote the portion of the
signal that corresponds to the first $\ell$ rounds of transmission.

Next, we describe the decoder. Since the performance analysis for
the optimal decoder seems intractable, in the sequel, we describe a
\emph{suboptimal} bounded distance decoder and analyze its
performance. Obviously, this analysis provides a \emph{lower bound}
on the diversity gain achieved through the protocol. To describe the
decoder, let us label the source and the receiving antenna that are
connected through the channel with the highest signal to
interference (due to the other source) and noise ratio by $s$ ($s$
stands for superior), while labeling the remaining source and
receiving antenna by $i$ ($i$ stands for inferior). This means that,
\begin{align}
\frac{|g_{ss}|^2\rho}{|g_{is}|^2\rho+\sigma^2} &\geq \max \{
\frac{|g_{si}|^2\rho}{|g_{ii}|^2\rho+\sigma^2},
\frac{|g_{is}|^2\rho}{|g_{ss}|^2\rho+\sigma^2},
\frac{|g_{ii}|^2\rho}{|g_{si}|^2\rho+\sigma^2}\}, \label{eq:44}
\end{align}
where, e.g., $g_{si}$ denotes the gain of the channel connecting
source $s$ to receive antenna $i$. Now, the decoder
$\{\mathbf{\varphi}, \mathbf{\psi}\}$, uses the two sets of
functions, $\mathbf{\varphi}=\{\varphi_j^1, \varphi_s^1,
\varphi_j^2, \varphi_i^2\}$ and $\mathbf{\psi}=\{\psi_j^1,
\psi_s^1\}$ to decode the messages, and produce the ACK/NACK
feedback bits, as follows:
\begin{enumerate}
\item At the end of the first round, the decoder uses $\varphi_j^1$
to \emph{jointly} decode the two messages, i.e.
\begin{align}
\varphi_j^1(\mathbf{Y}^1) &\triangleq \arg \min_{\mathbf{m}}
||\mathbf{Y}^1-\mathbf{S}^1(\mathbf{m})||^2.\label{eq:116}
\end{align}
We denote the event that $\varphi_j^1(\mathbf{Y}^1)$ is different
from the actual joint message sent, by $E_j^1$.

\item To decide whether it has correctly decoded the two messages
or not, the decoder uses $\psi_j^1$, where $\psi_j^1$ outputs a one,
if $\mathbf{m}$ is the \emph{unique} joint message satisfying
\begin{align}
||\mathbf{Y}^1 - \mathbf{S}^1(\mathbf{m})||^2 &\leq
2T(1+\delta)\sigma^2.\label{eq:117}
\end{align}
In \eqref{eq:117}, $\delta$ is some positive value. In any other
case, $\psi_j^1$ outputs a zero. Now, if $\psi_j^1(\mathbf{Y}^1)=1$,
then the decoder sends back ACK signals to \emph{both} of the users,
declaring $\varphi_j^1(\mathbf{Y}^1)$ as the decoded joint message.
This causes the two sources to start transmission of their next
messages. Otherwise, it proceeds to the next step as described
below. We denote the event $\psi_j^1(\mathbf{Y}^1)=1$, by $A_j^1$.

\item At the end of the first round and in the event of failure in jointly
decoding the two messages, i.e., $\overline{A_j^1}$, the decoder uses
$\varphi_s^1$ to decode the superior message, treating the inferior source's
contribution as interference. In doing so, the decoder only utilizes
the signal it has received through its $s$ antenna, i.e.
\begin{align}
\varphi_s^1(\mathbf{y}_s^1) &\triangleq \arg \min_{m_s}
|\mathbf{y}_s^1 - \mathbf{s}_s^1(m_s)|^2. \label{eq:46}
\end{align}
We denote the event that $\varphi_s^1(\mathbf{y}_s^1)$ is different
from the actual superior message sent, by $E_s^1$.

\item To decide whether it has correctly decoded the superior message
or not, the decoder uses $\psi_s^1$, where $\psi_s^1$ outputs a one,
if $m_s$ is the \emph{unique} superior message satisfying
\begin{align}
|\mathbf{y}_s^1 - \mathbf{s}_s^1(m_s)|^2 &\leq
T(1+\delta)(|g_{is}|^2\rho+\sigma^2).\label{eq:48}
\end{align}
In \eqref{eq:48}, $\delta$ is some positive value. In any other
case, $\psi_s^1$ outputs a zero. Now, if
$\psi_s^1(\mathbf{y}_s^1)=0$, the decoder requests for a second
round of transmission by sending back NACK signals to \emph{both} of
the sources. However, if $\psi_s^1(\mathbf{y}_s^1)=1$, then the
decoder sends back an ACK signal to the superior source, declaring
$\varphi_s^1(\mathbf{y}_s^1)$ as the decoded superior message, while
requesting a second round of transmission for the inferior message
by sending back a NACK signal to the inferior source. We denote the
event $\psi_s^1(\mathbf{y}_s^1)=1$, by $A_s^1$.

\item At the end of the second round and conditioned on successful
decoding of the superior message in the first round, i.e., $A_s^1
\cap \overline{A_j^1}$, the decoder declares
$\varphi_i^2(\mathbf{Y}^2,\varphi_s^1(\mathbf{y}_s^1))$ as the
decoded inferior message where
\begin{align}
\varphi_i^2(\mathbf{Y}^2,\varphi_s^1(\mathbf{y}_s^1)) &\triangleq
\arg \min_{m_i} ||\mathbf{Y}^2 -
\mathbf{S}^2(\varphi_s^1(\mathbf{y}_s^1)) -\mathbf{S}^2(m_i)
||^2.\label{eq:54}
\end{align}
In \eqref{eq:54}, $\varphi_s^1(\mathbf{y}_s^1)$ is the decoded
superior message as given by \eqref{eq:46}. We denote the event that
$\varphi_i^2(\mathbf{Y}^2,\varphi_s^1(\mathbf{y}_s^1))$ is different
from the actual inferior message sent, by $E_i^2$.

\item Finally, in case of failure in decoding the superior message at
the end of the first round, i.e., $\overline{A_s^1} \cap \overline{A_j^1}$,
the decoder declares $\varphi_j^2(\mathbf{Y}^2)$ as the decoded joint
message where
\begin{align}
\varphi_j^2(\mathbf{Y}^2) &\triangleq \arg \min_{\mathbf{m}}
||\mathbf{Y}^2-\mathbf{S}^2(\mathbf{m})||^2.\label{eq:67}
\end{align}
We denote the event that $\varphi_j^2(\mathbf{Y}^2)$ is different
from the actual joint message sent, by $E_j^2$.
\end{enumerate}

Having described the decoder, we next characterize its average error
probability $P_E$. Toward this end, we first notice that
\begin{align}
P_E &\leq P_{E|\overline{E_r}} + P_{E_r},\nonumber
\end{align}
where $E_r$ and $\overline{E_r}$ denote the event that the relaying
source makes an error in decoding the inferior message, and its
complement, respectively. Since the superior source only starts
relaying after the mutual information between its received signal
and inferior source's transmitted signal exceeds $R_1T/2$, we have
(e.g., refer to Theorem $10.1.1$ in \cite{CT:91})
\begin{align}
P_{E_r} &\leq \epsilon, \text{~~for any~~}\epsilon >0.\nonumber
\end{align}
This means that
\begin{align}
P_E &\dot{\leq} P_{E|\overline{E_r}}. \nonumber
\end{align}
For the sake of notational simplicity, in the sequel, we denote
$P_{E|\overline{E_r}}$ by $P_E$. To characterize $P_E$, we write
\begin{align}
P_E &= P_{E_j^1,A_j^1} + P_{E_s^1,A_s^1,\overline{A_j^1}} +
P_{E_i^2,\overline{E_s^1},A_s^1,\overline{A_j^1}} +
P_{E_j^2,\overline{A_s^1},\overline{A_j^1}}.\label{eq:68}
\end{align}
To understand \eqref{eq:68}, note that the first two terms
correspond to making an \emph{undetected} error in decoding one or
both of the messages at the end of the first round, while the last
two terms correspond to making a decoding error after requesting for
two rounds of transmission. Next, we upper bound \eqref{eq:68} by
\begin{align}
P_E &\leq P_{E_j^1,A_j^1} + P_{E_s^1,A_s^1} +
P_{E_i^2,\overline{E_s^1}} +
P_{E_j^2,\overline{A_s^1}}.\label{eq:69}
\end{align}
We start evaluating \eqref{eq:69} by characterizing $P_{E_j^1,A_j^1}$. By
examining the definitions for events $E_j^1$ and $A_j^1$ (refer to
\eqref{eq:116} and \eqref{eq:117}), and through an argument similar to the
one given for \eqref{eq:36}, we get
\begin{align}
P_{E_j^1,A_j^1|g,h} &\leq \Pr\{||\mathbf{N}^1||^2 >
2T(1+\delta)\sigma^2\},\nonumber
\end{align}
which for large enough $T$ gives (compare to \eqref{eq:55})
\begin{align}
P_{E_j^1,A_j^1|g,h} &\leq \epsilon\text{~~for any~~}\epsilon >
0,\text{~~or}\nonumber\\
P_{E_j^1,A_j^1} &\leq \epsilon\text{~~for any~~}\epsilon >
0.\label{eq:118}
\end{align}
Next, we characterize $P_{E_s^1,A_s^1}$. Toward this end, we first
fix a channel realization. Then, through examining the definitions
for events $E_s^1$ and $A_s^1$ (refer to \eqref{eq:46} and
\eqref{eq:48}), and by pursuing the same steps which led to
\eqref{eq:36}, we get
\begin{align}
P_{E_s^1,A_s^1|g,h} &\leq \Pr\{|\mathbf{s}_s^1(m_i)+\mathbf{n}_s^1|^2 >
T(1+\delta)(|g_{is}|^2\rho+\sigma^2)\},\nonumber
\end{align}
where $m_i$ denotes the actual inferior message sent and
$\mathbf{s}_s^1(m_i)$ represents its signature, at the end of the
first round, at the superior antenna. Realizing that, conditioned on
a certain channel realization,
$|\mathbf{s}_s^1(m_i)+\mathbf{n}_s^1|^2$ has a central Chi-squared
distribution with $2T$ degrees of freedom, we conclude that for
large enough $T$, we have (compare to \eqref{eq:55})
\begin{align}
P_{E_s^1,A_s^1|g,h} &\leq \epsilon\text{~~for any~~}\epsilon >
0,\text{~~or}\nonumber\\
P_{E_s^1,A_s^1} &\leq \epsilon\text{~~for any~~}\epsilon >
0.\label{eq:70}
\end{align}
In other words, \eqref{eq:118} and \eqref{eq:70} mean that, through
using long enough codes, one can make the probability of making
undetected errors arbitrarily small. Note that for doing so, the
bounded distance decoder does not employ any kind of cyclic
redundancy check (CRC) techniques. Now, using \eqref{eq:69},
\eqref{eq:118} and \eqref{eq:70} we conclude
\begin{align}
P_E &\dot{\leq} P_{E_i^2,\overline{E_s^1}} +
P_{E_j^2,\overline{A_s^1}}.\label{eq:72}
\end{align}
To characterize $P_{E_i^2,\overline{E_s^1}}$, we first
fix a channel realization and then write
\begin{align}
P_{E_i^2,\overline{E_s^1}|g,h} &=
P_{\overline{E_s^1}|g,h}
P_{E_i^2|\overline{E_s^1},g,h},\nonumber\\
&\leq P_{E_i^2|\overline{E_s^1},g,h}.\label{eq:86}
\end{align}
Now, using \eqref{eq:54}, it is straightforward to verify that
(refer to \cite{AES:05})
\begin{align}
P_{PE_i^2|\overline{E_s^1},g,h} \leq&
\big(1+\frac{1}{2}\rho(|g_{is}|^2+|g_{ii}|^2)\big)^{-(T+T')} \times
\nonumber \\
&\big(1+\frac{1}{2}\rho(|g_{ss}|^2+|g_{si}|^2+|g_{is}|^2+|g_{ii}|^2)
+\frac{1}{4}\rho^2\det(GG^H)\big)^{-(T-T')}, \label{eq:78}
\end{align}
where
\begin{align}
G &\triangleq \begin{bmatrix} g_{ss} & g_{is} \\ g_{si} &
g_{ii}\end{bmatrix}. \nonumber
\end{align}
In \eqref{eq:78}, $T'$ is the number of symbol intervals, in the
second round, that the superior source needs to listen to the
inferior one, before decoding its message, i.e.
\begin{align}
T' &\triangleq \min\{T,\lceil\frac{TR_1}{2\log_2 (1+
|h|^2c\rho)}\rceil\}.\label{eq:74}
\end{align}
Since $GG^H$ is a positive semi-definite matrix, we have
\begin{align}
P_{PE_i^2|\overline{E_s^1},g,h} \leq&
\big(1+\frac{1}{2}\rho(|g_{is}|^2+|g_{ii}|^2)\big)^{-(T+T')} \times
\nonumber \\
&\big(1+\frac{1}{2}\rho(|g_{ss}|^2+|g_{si}|^2+|g_{is}|^2+|g_{ii}|^2)\big)^{-(T-T')}.
\label{eq:73}
\end{align}
Note that in deriving \eqref{eq:73} from \eqref{eq:78}, we have
ignored the term $\frac{1}{4}\rho^2\det(GG^H)$. As a consequence,
the derived upper bound may be loose. However, this is indeed
necessary, for the sake of analysis tractability. Now, let us define
$v_{kl}$, where $k,l\in\{s,i\}$, as the exponential order of
$1/|g_{kl}|^2$, and $u$, as the exponential order of $1/|h|^2$.
Then, using \eqref{eq:73} and realizing that there are a total of
$\rho^{\frac{Tr_1}{2}}$ codewords in the inferior source's
code-book, we derive
\begin{align}
P_{E_i^2|\overline{E_s^1},v,u} \dot{\leq}&
\rho^{-T[(1-\min\{v_{is},v_{ii}\})^+(1+f)+
(1-\min\{v_{ss},v_{si},v_{is},v_{ii}\})^+(1-f)-\frac{r_1}{2}]},
\label{eq:75}
\end{align}
where $f \triangleq T'/T$. Using \eqref{eq:74}, we have
\begin{align}
f &= \min\{1,\frac{r_1}{2(1-u)^+}\}.\label{eq:76}
\end{align}
An argument similar to that given for \eqref{eq:18}, reveals that
if we define the outage event $O_i^{2+}$ as
\begin{align}
O_i^{2+} \triangleq \{&(v_{ss},\cdots,u) \in \Real^{5+}|
(1-\min\{v_{is},v_{ii}\})^+(1+f)+ \nonumber \\
&(1-\min\{v_{ss},v_{si},v_{is},v_{ii}\})^+(1-f) \leq
\frac{r_1}{2}\},\label{eq:77}
\end{align}
then
\begin{align}
P_{E_i^2|\overline{E_s^1}}~&\dot{\leq}~
P_{O_i^2},\label{eq:79}
\end{align}
where (recall \eqref{eq:23})
\begin{align}
P_{O_i^2} &\dot{=}
\rho^{-d_i(r_1)},\text{~~with~~} d_i(r_1)
\triangleq \inf_{O_i^{2+}}
\{v_{ss}+v_{si}+v_{is}+v_{ii}+u\}.\label{eq:80}
\end{align}
In writing \eqref{eq:80}, we have used the fact that
\begin{align}
v_{ss}+v_{si}+v_{is}+v_{ii} &= v_{11}+v_{12}+v_{21}+v_{22}.\nonumber
\end{align}
Obviously, $(v_{ss},v_{si},v_{is},v_{ii})$ should also satisfy
\begin{align}
1-v_{ss}-(1-v_{is})^+ \geq \max\{&1-v_{si}-(1-v_{ii})^+,
1-v_{is}-(1-v_{ss})^+,\nonumber\\
&1-v_{ii}-(1-v_{si})^+\},\label{eq:82}
\end{align}
which is the counterpart of \eqref{eq:44}, stated in terms of
the channel exponential orders. Now, to characterize
$d_i(r_1)$, we first derive $\inf_{O_i^{2+}}
\{v_{ss}+v_{si}+v_{is}+v_{ii}\}$, as a function of $f$, i.e.
\begin{align}
\inf_{O_i^{2+}} \{v_{ss}+v_{si}+v_{is}+v_{ii}\}(f) &=
\lambda_i(f),\nonumber
\end{align}
where
\begin{align}
\lambda_i(f) &\triangleq \left\{
\begin{array}{ll}
4-\frac{r_1}{1-f}, & 1-\frac{r_1}{2} > f \ge 0\\
\frac{4-r_1}{1+f}, & 1 \geq f \ge 1-\frac{r_1}{2}
\end{array}\label{eq:81}
\right. .
\end{align}
On the other hand, from \eqref{eq:76} we get
\begin{align}
\inf \{u\}(f) &= \lambda(f),\text{~~where~~} \lambda(f)
\triangleq 1-\frac{r_1}{2f},\text{~~}f \geq \frac{r_1}{2}.
\label{eq:83}
\end{align}
Then, from \eqref{eq:80}, we get
\begin{align}
d_i(r_1) &= \inf_{1 \geq f \geq \frac{r_1}{2}}
\lambda_i(f) + \lambda(f).\nonumber
\end{align}
The right hand side of this expression can be derived using \eqref{eq:81}
and \eqref{eq:83}, i.e.
\begin{align}
d_i(r_1) &= \left\{
\begin{array}{ll}
3-r_1, & 1 > r_1 \ge 0\\
\frac{2(4-r_1)}{2+r_1}, & 2 \geq r_1 \ge 1
\end{array}\label{eq:84}
\right..
\end{align}
This, together with \eqref{eq:80}, \eqref{eq:79} and \eqref{eq:86},
completes the characterization of $P_{E_i^2,\overline{E_s^1}}$,
i.e.
\begin{align}
P_{E_i^2,\overline{E_s^1}}~&\dot{\leq}~
\rho^{-d_i(r_1)}.\label{eq:85}
\end{align}
To characterize the second term of \eqref{eq:72}, i.e.,
$P_{E_j^2,\overline{A_s^1}}$, we first split the event
$E_j^2$ (refer to \eqref{eq:67}) into
\begin{align}
E_j^2 &= E_{jsi}^2 \cup E_{ji}^2 \cup
E_{js}^2, \nonumber
\end{align}
where $E_{jsi}^2$, $E_{ji}^2$ and $E_{js}^2$ represent the events that,
at the end of the second round, the joint decoder makes errors in decoding
both of the messages, only the inferior message and only the superior
message, respectively. Since these three events are mutually exclusive, we
have
\begin{align}
P_{E_j^2,\overline{A_s^1}} &=
P_{E_{jsi}^2,\overline{A_s^1}} +
P_{E_{ji}^2,\overline{A_s^1}} +
P_{E_{js}^2,\overline{A_s^1}}.\label{eq:87}
\end{align}
Characterization of the first term is very straightforward,
\begin{align}
P_{E_{jsi}^2,\overline{A_s^1}} &\leq
P_{E_{jsi}^2},\nonumber\\
&\dot{=} \rho^{-d_{2 \times 2}(\frac{r_1}{2})}.\label{eq:88}
\end{align}
Characterization of $P_{E_{ji}^2,\overline{A_s^1}}$, though, is a
little bit more involved. In particular, notice that
\begin{align}
P_{E_{ji}^2,\overline{A_s^1}} &\leq
\min\{P_{E_{ji}^2},P_{\overline{A_s^1}}\}.\label{eq:89}
\end{align}
Now, conditioned on a certain channel realization,
$P_{PE_{ji}^2|g,h}$ can be upper bounded as (refer to
\cite{AES:05})
\begin{align}
P_{PE_{ji}^2|g,h} &\leq
\big(1+\frac{1}{2}\rho(|g_{is}|^2+|g_{ii}|^2)\big)^{-2T}.\nonumber
\end{align}
Realizing that there are a total of $\rho^{\frac{Tr_1}{2}}$ pairs of
codewords in each source's code book, we get
\begin{align}
P_{E_{ji}^2|v,u} &\dot{\leq}
\rho^{-2T[(1-\min\{v_{is},v_{ii}\})^+-\frac{r_1}{4}]}.\label{eq:90}
\end{align}
Now, examining \eqref{eq:90} reveals that if we define $O_{ji}^{2+}$ as
\begin{align}
O_{ji}^{2+} &\triangleq \{(v_{ss},\cdots,u) \in \Real^{5+}|
(1-\min\{v_{is},v_{ii}\})^+ \leq \frac{r_1}{4}\}, \label{eq:91}
\end{align}
then for all channel realizations $(v_{ss},\cdots,u) \in \Real^{5+}$ \emph{not}
included in $O_{ji}^{2+}$,  $P_{E_{ji}^2|v,u}$ can be made
arbitrary small, provided that $T$ is large enough, i.e.
\begin{align}
P_{E_{ji}^2|v,u} &\leq \epsilon,\text{~~for any~~}\epsilon
>0\text{~~and~~} (v_{ss},\cdots,u) \in \overline{O_{ji}^2} \cap
\Real^{5+}. \label{eq:92}
\end{align}
Next, we turn our attention to $P_{\overline{A_s^1}}$. To characterize this
probability, we first fix a channel realization. Then, by comparing the
definition of $A_s^1$ (refer to \eqref{eq:48}) to that of $A^{\ell}$ (refer to
\eqref{eq:37}), and pursuing the exact same arguments which led \eqref{eq:93}
to \eqref{eq:59}, we conclude
\begin{align}
P_{\overline{A_s^1}|g,h} ~&\dot{\leq}~ P_{E_s^1|g,h}.\label{eq:94}
\end{align}
But $P_{PE_s^1|g,h}$ is given by (notice how the inferior source's
contribution is treated as interference),
\begin{align}
P_{PE_s^1|g,h} &\leq
\left(1+\frac{1}{2} \rho
(\frac{|g_{ss}|^2}{1+\rho|g_{is}|^2})\right)^{-T},\nonumber
\end{align}
which, in terms of channel exponentials, translates into
\begin{align}
P_{E_s^1|v,u} &\dot{\leq}
\rho^{-T\big[\big(1-v_{ss}-(1-v_{is})^+\big)^+-\frac{r_1}{2}\big]}.\label{eq:95}
\end{align}
Now, from \eqref{eq:94} and \eqref{eq:95}, we conclude
\begin{align}
P_{\overline{A_s^1}|v,u} ~&\dot{\leq}
\rho^{-T\big[\big(1-v_{ss}-(1-v_{is})^+\big)^+-\frac{r_1}{2}\big]}.\label{eq:96}
\end{align}
This means that if
\begin{align}
O_s^{1+} &\triangleq \{(v_{ss},\cdots,u) \in \Real^{5+} |
\big(1-v_{ss}-(1-v_{is})^+\big)^+ \leq \frac{r_1}{2}\},\label{eq:98}
\end{align}
then for all channel realizations $(v_{ss},\cdots,u) \in
\Real^{5+}$, \emph{not} included in $O_s^{1+}$,
$P_{\overline{A_s^1}|v,u}$ can be made arbitrary small, provided that
$T$ is large enough, i.e.
\begin{align}
P_{\overline{A_s^1}|v,u} &\leq \epsilon,\text{~~for all~~}\epsilon
>0\text{~~and~~} (v_{ss},\cdots,u) \in \overline{O_s^1} \cap
\Real^{5+}.\label{eq:97}
\end{align}
Now, using \eqref{eq:89}, \eqref{eq:92} and \eqref{eq:97} we conclude
\begin{align}
P_{E_{ji}^2,\overline{A_s^1}|u,v} &\leq \epsilon \text{~~for
  all~~}\epsilon >0\text{~~and~~}(v_{ss},\cdots,u) \in
  \overline{O_{s,ji}^{1,2}} \cap \Real^{5+},\label{eq:99}
\end{align}
where
\begin{align}
O_{s,ji}^{1,2+} &\triangleq O_s^{1+} \cap
O_{ji}^{2+},\label{eq:100}
\end{align}
or
\begin{align}
O_{s,ji}^{1,2+} = \{(v_{ss},\cdots,u) \in
\Real^{5+}| (1-\min\{v_{is},v_{ii}\})^+ \leq \frac{r_1}{4}&,\nonumber \\
\big(1-v_{ss}-(1-v_{is})^+\big)^+ \leq \frac{r_1}{2}&\}.\label{eq:103}
\end{align}
This means that
\begin{align}
P_{E_{ji}^2,\overline{A_s^1}} &\dot{\leq}
P_{O_{s,ji}^{1,2}},\label{eq:104}
\end{align}
where
\begin{align}
P_{O_{s,ji}^{1,2}} &\dot{=}
\rho^{-d_{s,ji}(r_1)},\text{~~with~~}
d_{s,ji}(r_1) \triangleq
\inf_{O_{s,ji}^{12+}}
\{v_{ss}+\cdots+u\}.\label{eq:101}
\end{align}
Now, using \eqref{eq:103}, it is straightforward to show that
\begin{align}
d_{s,ji}(r_1) &= \left\{
\begin{array}{ll}
4-2r_1, & \frac{4}{3} > r_1 \ge 0\\
2-\frac{r_1}{2}, & 2 \geq r_1 \ge \frac{4}{3}
\end{array}\label{eq:102}
\right..
\end{align}
This, together with \eqref{eq:101} and \eqref{eq:104}, completes the
characterization of $P_{E_{ji}^2,\overline{A_s^1}}$, i.e.
\begin{align}
P_{E_{ji}^2,\overline{A_s^1}} &\dot{\leq}
\rho^{-d_{s,ji}(r_1)}.\label{eq:105}
\end{align}
Characterization of $P_{E_{js}^2,\overline{A_s^1}}$, proceeds in
a similar way. In particular, pursuing the exact same steps leading to
\eqref{eq:90}, reveals that
\begin{align}
P_{E_{js}^2|v,u} &\dot{\leq}
\rho^{-2T[(1-\min\{v_{ss},v_{si}\})^+-\frac{r_1}{4}]}.\nonumber
\end{align}
This means that defining $O_{js}^{2+}$ as
\begin{align}
O_{js}^{2+} &\triangleq \{(v_{ss},\cdots,u) \in \Real^{5+}|
(1-\min\{v_{ss},v_{si}\})^+ \leq \frac{r_1}{4}\}, \nonumber
\end{align}
results in
\begin{align}
P_{E_{js}^2|v,u} &\leq \epsilon,\text{~~for any~~}\epsilon
>0\text{~~and~~} (v_{ss},\cdots,u) \in \overline{O_{js}^2} \cap
\Real^{5+}. \nonumber
\end{align}
This, however, together with \eqref{eq:97}, gives
\begin{align}
P_{E_{js}^2,\overline{A_s^1}|u,v} &\leq \epsilon \text{~~for
  all~~}\epsilon >0\text{~~and~~}(v_{ss},\cdots,u) \in
  \overline{O_{s,js}^{1,2}} \cap \Real^{5+},\nonumber
\end{align}
where
\begin{align}
O_{s,js}^{1,2+} &\triangleq O_s^{1+} \cap O_{js}^{2+}.\nonumber
\end{align}
or
\begin{align}
O_{s,js}^{1,2+} = \{(v_{ss},\cdots,u) \in
\Real^{5+}| (1-\min\{v_{ss},v_{si}\})^+ \leq \frac{r_1}{4},&\nonumber \\
\big(1-v_{ss}-(1-v_{is})^+\big)^+ \leq \frac{r_1}{2}\}&,\nonumber
\end{align}
or
\begin{align}
O_{s,js}^{1,2+} = \{(v_{ss},\cdots,u) \in
\Real^{5+}| (1-\min\{v_{ss},v_{si}\})^+ \leq \frac{r_1}{4}\}.\label{eq:106}
\end{align}
Thus
\begin{align}
P_{E_{js}^2,\overline{A_s^1}} &\dot{\leq}
\rho^{-d_{s,js}(r_1)},\label{eq:107}
\end{align}
where
\begin{align}
d_{s,js}(r_1) \triangleq
\inf_{O_{s,js}^{1,2+}}
\{v_{ss}+\cdots+u\}.\nonumber
\end{align}
Now using  \eqref{eq:106}, together with \eqref{eq:82}, it is a simple matter
to show that
\begin{align}
d_{s,js}(r_1) &=4-r_1.\label{eq:108}
\end{align}
This completes the characterization of
$P_{E_{js}^2,\overline{A_s^1}}$. Next, we use \eqref{eq:87},
\eqref{eq:88}, \eqref{eq:105} and \eqref{eq:107} we conclude
\begin{align}
P_{E_j^2,\overline{A_s^1}} &\dot{\leq}
\rho^{-d_{s,j}(r_1)},\label{eq:109}
\end{align}
where
\begin{align}
d_{s,j}(r_1) &= \min\{d_{2 \times 2}(r_1), d_{s,ji}(r_1),
d_{s,js}(r_1)\}.\nonumber
\end{align}
Using \eqref{eq:102} and \eqref{eq:108}, however, we get
\begin{align}
d_{s,j}(r_1) &= d_{s,ji}(r_1).\label{eq:110}
\end{align}
Finally, \eqref{eq:72}, together with \eqref{eq:85} and \eqref{eq:109},
gives
\begin{align}
d_{\text{DDF-CVMA}}(r_1,2) &\geq \min\{d_i(r_1),
d_{s,j}(r_1)\},\nonumber
\end{align}
where $d_{\text{DDF-CVMA}}(r_1,2)$ denotes the diversity gain
achieved by the protocol. Now, using \eqref{eq:84}, \eqref{eq:110}
and \eqref{eq:102} we get
\begin{align}
d_{\text{DDF-CVMA}}(r_1,2) &\ge \left\{ \begin{array}{ll}
3-r_1, & 1 > r_1 \ge 0 \\
4-2r_1, & \frac{4}{3} > r_1 \ge 1 \\
2-\frac{r_1}{2}, & 2 > r_1 \ge \frac{4}{3}
\end{array}\label{eq:111}
\right.
\end{align}
Next, we prove that for $2 > r_1 \geq 0$, $r_{\text{e}}=r_1$. To do
this, we only need to characterize $p(1)$ (recall \eqref{eq:39}).
Toward this end, we observe that
\begin{align}
p(1) =& P_{\overline{A_j^1}},\nonumber\\
\dot{=}& P_{E_j^1}.\label{eq:114}
\end{align}
$P_{E_j^1}$, however, is the joint error probability, for a multiple
access channel with two single-antenna users and one double-antenna
destination, which is known to be (refer to \cite{TVZ:03})
\begin{align}
P_{E_j^1} &\dot{=} \rho^{-(2-r_1)}.\label{eq:115}
\end{align}
Now, \eqref{eq:115}, together with \eqref{eq:114}, proves that
$p(1)$ decays polynomially with $\rho$ over the range $2 > r_1 \geq
0$. Thus
\begin{align}
r_e &= r_1 \text{~~for~~} 2 > r_1 \geq 0.\label{eq:119}
\end{align}
From \eqref{eq:119} and \eqref{eq:111}, we conclude
\begin{align}
d_{\text{DDF-CVMA}}(r_e,2) &\ge \left\{ \begin{array}{ll}
3-r_e, & 1 > r_e \ge 0 \\
4-2r_e, & \frac{4}{3} > r_e \ge 1 \\
2-\frac{r_e}{2}, & 2 > r_e \ge \frac{4}{3}
\end{array}\label{eq:120}
\right.
\end{align}
Now, application of Lemma $11$ of \cite{ECD:04} shows that there are
codes in the ensemble that achieve \eqref{eq:114} and
\eqref{eq:120}, \emph{simultaneously}. This proves the achievability
of \eqref{eq:12}.

Next, we prove the asymptotic optimality result. To do this,
however, we first need to generalize the protocol. To extend the
protocol to the case of $L$ ARQ rounds, the random codes
$C(\rho)=\{C_1(\rho),C_2(\rho)\}$ used by the two sources should be
modified such that each one is of length $LT$ symbols and rate
$R_1/2L$ BPCU. Notice that these choices leave the total number of
information bits per joint message equal to $R_1T$ bits. The decoder
used by the destination is also a direct extension of the one
described in the proof of the achievability part. In particular, at
the end of the $\ell$th round ($L \geq \ell \geq 1$) and in the
event that none of the two messages has yet been successfully
decoded, the decoder uses $\varphi_j^{\ell}$ and $\psi_j^{\ell}$ to
jointly decode the two messages ($\varphi_j^{\ell}$ and
$\psi_j^{\ell}$ are defined in a way similar to \eqref{eq:116} and
\eqref{eq:117}). If successful, the decoder sends back ACK signals
to both of the users (i.e., event $A_j^{\ell}$), otherwise it tries
to decode the superior message, using $\varphi_s^{\ell}$ and
$\psi_s^{\ell}$ (which are defined in a way similar to \eqref{eq:46}
and \eqref{eq:48}). In the case that the decoder succeeds in
decoding the superior message (i.e., event $A_s^{\ell}$), an ACK
signal is sent to the corresponding user. The description given so
far pertains to the scenario where none of the messages is
successfully decoded by the end of the $(\ell -1)$th round. If the
superior message is already decoded, the decoder uses
$\varphi_i^{\ell}$ ($L \geq \ell > 1$) to decode the inferior
message, i.e.
\begin{align}
\varphi_i^{\ell}(\mathbf{Y}^{\ell}, m_s) &\triangleq \arg \min_{m_i}
||\mathbf{Y}^{\ell} - \mathbf{S}^{\ell}(m_s) -\mathbf{S}^{\ell}(m_i)
||^2, \label{eq:122}
\end{align}
where $m_s$ denotes the decoded superior message. To decide whether
it has decoded the inferior message error-free or not, the decoder
uses $\psi_i^{\ell}$, where $\psi_i^{\ell}$ outputs a one, if $m_i$
is the \emph{unique} message satisfying
\begin{align}
||\mathbf{Y}^{\ell} - \mathbf{S}^{\ell}(m_s) -\mathbf{S}^{\ell}(m_i)
||^2 &\leq 2 \ell T (1+\delta) \sigma^2. \nonumber
\end{align}
In any other case, $\psi_i^{\ell}$ outputs a zero. Now, if
$\psi_i^{\ell}(\mathbf{Y}^{\ell},m_s)=1$, then the decoder sends
back an ACK signal to the inferior user (i.e., event $A_i^{\ell}$),
declaring $\varphi_i^{\ell}(\mathbf{Y}^{\ell},m_s)$ as the decoded
inferior message.

Having described the decoder, we next characterize its error
probability, $P_E$, as $L$ grows to infinity. Note that $P_E$ can be
written as
\begin{align}
P_E &= \Sigma_{\ell=1}^L P_{E^{\ell}}, \label{eq:123}
\end{align}
where $E^{\ell}$ denotes the event of making an error at the end of
the $\ell$th round. It is important to realize that for $\ell < L$,
$E^{\ell}$ corresponds to an \emph{undetected} error, i.e., one for
which an ACK signal is sent back. Now, an argument similar to those
given for \eqref{eq:55}, \eqref{eq:118} and \eqref{eq:70} reveals
that for large enough $T$,
\begin{align}
P_{E^{\ell}} &\leq \epsilon,\text{~~for~~} L > \ell \geq 1.\nonumber
\end{align}
This together with \eqref{eq:123} gives
\begin{align}
P_E &\dot{\leq} P_{E^L}. \label{eq:124}
\end{align}
But
\begin{align}
P_{E^L} &\leq \Sigma_{\ell=1}^{L-1}
P_{E_i^L,\overline{E_s^{\ell}},A_s^{\ell},\overline{A_j^{\ell}}} +
P_{E_j^L,\overline{A_s^{L-1}},\overline{A_j^{L-1}}}, \label{eq:125}
\end{align}
where
$P_{E_i^L,\overline{E_s^{\ell}},A_s^{\ell},\overline{A_j^{\ell}}}$
upper-bounds the probability that the decoder makes an error after
$L$ rounds of transmission, conditioned on successful decoding of
the superior message at the end of the $\ell$th round.
$P_{E_j^L,\overline{A_s^{L-1}},\overline{A_j^{L-1}}}$, on the other
hand, upper-bounds the decoder error probability after $L$ rounds,
given that it does not decode any of the messages by the $(L-1)$th
round. To characterize
$P_{E_i^L,\overline{E_s^{\ell}},A_s^{\ell},\overline{A_j^{\ell}}}$,
notice that
\begin{align}
P_{E_i^L,\overline{E_s^{\ell}},A_s^{\ell},\overline{A_j^{\ell}}}
&\leq \min
\{P_{E_i^L,\overline{E_s^{\ell}},A_s^{\ell}},P_{\overline{A_j^{\ell}}}\},
\text{~~for~~} L > \ell \geq 1.\label{eq:126}
\end{align}
Since we are only interested in the case where $L$ grows to
infinity, we can assume that $L$ is even. Now for $\frac{L}{2} \geq
\ell \geq 1$, we have
\begin{align}
P_{E_i^L,\overline{E_s^{\ell}},A_s^{\ell}} &\leq
P_{E_i^L,\overline{E_s^{L/2}},A_s^{L/2}}.\label{eq:127}
\end{align}
This is because in the event $E_i^L\overline{E_s^{\ell}}A_s^{\ell}$,
$\ell$ corresponds to the number of rounds during which the superior
user is acting as a DDF relay for the inferior one. Likewise, for
$\ell \geq \frac{L}{2}+1$
\begin{align}
P_{\overline{A_j^{\ell}}} &\leq
P_{\overline{A_j^{L/2}}}.\label{eq:128}
\end{align}
This is because in the event $\overline{A_j^{\ell}}$, $\ell$
corresponds to the number of rounds during which the two users are
simultaneously transmitting their messages. Clearly, increasing
$\ell$ reduces $P_{\overline{A_j^{\ell}}}$ (notice that for exactly
the same reason, $P_{E_j^L} \leq P_{E_j^{L/2}}$). Following the same
steps leading to \eqref{eq:59} reveals that
\begin{align}
P_{\overline{A_j^{L/2}}} &\dot{\leq} P_{E_j^{L/2}},\label{eq:129}
\end{align}
which simply states that the probability of sending back a NACK
signal is of the same exponential order as the probability of making
a joint error. Now from equations \eqref{eq:125} to \eqref{eq:129},
we conclude
\begin{align}
P_{E^L} &\dot{\leq} \frac{L}{2}
(P_{E_i^L,\overline{E_s^{L/2}},A_s^{L/2}} +
P_{\overline{E_j^{L/2}}}),\text{~~for $L$ even}.\label{eq:130}
\end{align}
Comparing \eqref{eq:130} with \eqref{eq:72}, we realize that
$P_{E^L}$ is of same exponential order as the error probability of
the ARQ-DDF CVMA protocol, with two rounds of transmission and a
destination first-round rate of $\frac{2R_1}{L}$ BPCU, i.e.
\begin{align}
P_{E^L} &\dot{\leq}
\rho^{-d_{\text{DDF-CVMA}}(\frac{2r_1}{L},2)},\text{~~for~~} 2 > r_1
\geq 0, \nonumber
\end{align}
where $d_{\text{DDF-CVMA}}(r_1,2)$ is given by \eqref{eq:111}. Now,
letting $L$ to grow to infinity, together with \eqref{eq:124} and
\eqref{eq:111}, gives
\begin{align}
\lim_{L \to \infty} P_E &\dot{\leq} \rho^{-3},\text{~~for~~} 2
> r_1 \geq 0.\label{eq:132}
\end{align}
To prove that $r_e = r_1$, we notice that (recall \eqref{eq:39})
\begin{align}
\eta &\leq \frac{R_1}{1+p(1)},\nonumber
\end{align}
where
\begin{align}
p(1) =&  P_{\overline{A_j^1}},\nonumber\\
\dot{=}& \rho^{-(2-r_1)}.\label{eq:131}
\end{align}
The last step of \eqref{eq:131} follows from the same argument given
for \eqref{eq:115}. This shows that $r_e = r_1, 2 > r_1 \geq 0$,
which together with \eqref{eq:132} proves \eqref{eq:121}. Note that
the proof for existence of codes that achieve $P_E$ and $\eta$
simultaneously results from application of Lemma $11$ of
\cite{ECD:04}.


\end{document}